\documentclass{CVM}

\usepackage[nolist]{acronym}
\usepackage{wrapfig,lipsum,booktabs}
\usepackage{anyfontsize}

\CVMsetup{
type      = {Research Article},
doi       = {s41095-0xx-xxxx-x},
title     = {A Multi-scale Yarn Appearance Model with Fiber Details},
author    = {Apoorv Khattar$^{1}$ (Corresponding Author), Junqiu Zhu$^{2}$, Jean-Marie Aubry$^{3}$, Emiliano Padovani$^{3}$, Marc Droske$^{3}$, Ling-Qi Yan$^{2}$, Zahra Montazeri$^{1}$},
runauthor = {A. Khattar, J. Zhu, J.M. Aubry et al.},
abstract  = {
    Rendering  cloth realistically has always been a challenge due to its intricate structure. Cloth is made up of fibers, plies, and yarns, and previous curve-based models, while detailed, were computationally expensive and inflexible for large pieces of cloth. To address this, we propose a simplified approach. We introduce a geometric aggregation technique that reduces ray-tracing computation by using fewer curves, focusing only on yarn curves. Our model generates ply and fiber shapes implicitly, compensating for the lack of explicit geometry with a novel shadowing component. We also present a shading model that simplifies light interactions between fibers by categorizing them into four components, accurately capturing specular and scattered light in both forward and backward directions. To render large pieces of cloth efficiently, we propose a multi-scale solution based on pixel coverage. Our yarn shading model can be rendered 3-5 times faster   with less memory, for near-field views, compared to fiber-based models. Additionally, our multi-scale solution offers a 20\% speed boost for distant cloth observation.
},
keywords  = {Cloth, Rendering, Yarn, Shading, Bidirectional Scattering Distribution Function (BSDF)},
copyright = {The Author(s) 2024},
}

\begin{document}

\maketitle

\begin{figure}[b] 
    \small\renewcommand\arraystretch{1.3}
    \begin{tabular}{p{80.5mm}} \toprule\\ \end{tabular}
    \vskip -4.5mm \noindent \setlength{\tabcolsep}{1pt}
    \begin{tabular}{p{3.5mm}p{80mm}}
    $1\quad $ & The University of Manchester, Manchester, M13 9PL, UK. E-mail: A. Khattar apoorv.khattar@manchester.ac.uk (Corresponding Author); Z. Montazeri: zahra.montazeri@manchester.ac.uk \\
    $2\quad $ & University of California Santa Barbara, California, United States of America. Email: J. Zhu: zhujunqiu@mail.sdu.edu.cn; L. Yan: lingqi@cs.ucsb.edu \\
    $3\quad $ & WETA FX, Wellington, New Zealand. Email: J.M. Aubry: jaubry@wetafx.co.nz; E. Padovani: emilianop@wetafx.co.nz; M. Droske: mdroske@wetafx.co.nz\\
    \end{tabular} \vspace {-3mm}
\end{figure}

\begin{figure*}[t]
    \centering
    \setlength{\fboxsep}{0pt}
    \setlength{\fboxrule}{1.25pt}
    \includegraphics[width=0.99\linewidth]{teaser_error.pdf}
    %
    \caption{\label{fig_teaser}
     In this scene we compare the rendering results of our BYSDF model (near-field and multi-scale) to the reference ply-based model for a  knitted beanie, with  error maps provided in the insets. Our model achieves very similar results to the reference, both in distant and close-up views, with accurate soft shadows and geometry of plies and fibers as can be seen in the  close-ups, in top and bottom rows respectively. Our approach achieves these results while utilizing only 20\% of the memory at 2.5 times the speed of the reference rendering for the same  quality (noise-level). Our multi-scale result offers additional performance gain by leveraging a level-of-detail strategy as shown in the distant results. The scene was lit using two area lights and one constant environment lighting.
    }
    %
\end{figure*}

\newcommand{\important}[1]{\textcolor{olive}{\textbf{#1}}}

\begin{acronym}
\acro{BCSDF}{Bidirectional Curve Scattering Distribution Function}
\acro{BSDF}{Bidirectional Scattering Distribution Function}
\acro{BRDF}{Bidirectional Reflection Distribution Function}
\acro{BTF}{Bidirectional Texture Functions}
\acro{PDF}{Probability Density Function}
\acro{DS}{dual scattering}
\end{acronym}

\newcommand{\mypara}[1]{\par\textbf{#1:}\quad}

\newcommand*{\tabref}[1]{\tablename~\ref{#1}}
\newcommand*{\figref}[1]{Fig.~\ref{#1}}
\newcommand*{\algoref}[1]{Alg.~\ref{#1}}
\newcommand*{\equationref}[1]{Eqn.~\ref{#1}}
\newcommand*{\tableref}[1]{Table~\ref{#1}}

\newcommand{\mymath}[2]{
    \newcommand{#1}{\TextOrMath{$#2$\xspace}{#2}}}
\mymath{\azimuthalOffset}{h}
\newcommand{\revadded}[1]{#1}
\mymath{\azimuthalPhase}{\mathrm{u}}
\mymath{\longitudinalLength}{\mathrm{v}}

\mymath{\reflectionLobe}{\bsdfx^\mathrm{S}}
\mymath{\transmissionLobe}{\bsdfy^\mathrm{S}}
\mymath{\forwardScatteringLobe}{\bsdfx^\mathrm{B}}
\mymath{\backwardScatteringLobe}{\bsdfy^\mathrm{B}}

\newcommand{\fresnel}{F}
\newcommand{\fresnelr}{\fresnel_\mathrm{r}}
\newcommand{\fresnelt}{\fresnel_\mathrm{t}}
\newcommand{\fresneltDiffuse}{\fresnel_\mathrm{dr}}
\newcommand{\fiberShadow}{\mathrm{G_f}}
\newcommand{\plyShadow}{\mathrm{G_p}}
\newcommand{\albedo}{\mathrm{\alpha}}
\newcommand{\roughness}{\mathrm{\beta}}

\newcommand{\fspec}{\bsdf_\mathrm{specular}}
\newcommand{\fspecR}{\bsdf_\mathrm{specular}^\mathrm{R}}
\newcommand{\fspecTT}{\bsdf_\mathrm{specular}^\mathrm{TT}}
\newcommand{\fdiffR}{\bsdf_\mathrm{diffuse}^\mathrm{R}}
\newcommand{\fLdiff}{\bsdf_\mathrm{diffuse}^\mathrm{L}}
\newcommand{\fLSdiff}{\bsdf_\mathrm{diffuse}^\mathrm{LS}}
\newcommand{\fdiffTT}{\bsdf_\mathrm{diffuse}^\mathrm{TT}}
\newcommand{\ftrans}{\bsdf_\mathrm{trans}}
\newcommand{\fdtrans}{\bsdf_\mathrm{difftrans}}
\newcommand{\ATT}{\mathrm{A_{TT}}}
\newcommand{\abs}[1]{| #1 |}

\newcommand{\bx}{\bm{x}}
\newcommand{\by}{\bm{y}}

\newcommand{\bom}{\bm{\omega}}
\newcommand{\bomi}{\bom_\mathrm{i}}
\newcommand{\bomo}{\bom_\mathrm{o}}
\newcommand{\bomh}{\bom_\mathrm{h}}
\newcommand{\bomt}{\bom_\mathrm{t}}

\newcommand{\bomix}{\bom_{\bx}^\mathrm{i}}
\newcommand{\bomox}{\bom_{\bx}^\mathrm{o}}
\newcommand{\bomiy}{\bom_{\by}^\mathrm{i}}
\newcommand{\bomoy}{\bom_{\by}^\mathrm{o}}

\newcommand{\bsdf}{f}
\newcommand{\bsdfx}{\bsdf_{\bx}}
\newcommand{\bsdfy}{\bsdf_{\by}}
\newcommand{\bsdfnear}{\bsdf_\mathrm{near}}
\newcommand{\bsdffar}{\bsdf_\mathrm{far}}
\newcommand{\bsdfmulti}{\bsdf_\mathrm{multi}}

\newcommand{\Real}{\mathbb{R}}
\newcommand{\D}{\mathrm{d}}
\newcommand{\E}{\mathrm{e}}
\newcommand{\gnorm}{\bm{n}}
\newcommand{\sigT}{\sigma_\mathrm{t}}

\newcommand{\plyNum}{\bm{N}}
\newcommand{\plyNumIntersected}{\bm{N}_{i}}

\section{Introduction}
\label{sec_intro}

Fabrics are an integral part of our everyday lives, serving a wide range of purposes from clothing to functional textiles such as curtains, furniture fabrics, and tablecloths. Accurately modeling the appearance of cloth in a physically faithful manner has extensive applications in various fields, including design, online retail, and entertainment. However, fabrics present a challenge due to their complex geometry and optics, with a hierarchical structure consisting of yarns, plies, and individual fibers.

There are  two predominant approaches for cloth rendering: surface-based and curve-based methods. Surface-based models represent cloth geometry using 2D sheets, often in the form of polygonal meshes \cite{adabala2003real, sadeghi2013practical, irawan2012specular,zhu2022practical}. While these models offer lightweight and editable representations suitable for macroscopic views, they lack the fine-grained details required to convincingly render cloth in close-up views.

On the other hand, curve-based models focus on capturing intricate cloth details by representing the geometry at a microscale \cite{khungurn2015matching, zhao2011building}. These fiber-based models represent the structure of individual fibers and utilize volumetric or fiber-based light scattering models to simulate light interactions. While these approaches can achieve highly realistic renderings with exceptional fidelity, they suffer from computational expense, making them slow and challenging to manipulate.

To address these limitations, Montazeri et al. \cite{montazeri2020practical}  introduced a ply-based approach (referred to as the \emph{ply-based model}) that replaces the explicit representation of individual fibers with ply curves, implicitly incorporating fiber details. However, the number of plies in a yarn can vary (typically from 3 to 12) which makes modeling fabrics at the ply-level challenging and it still remains  computationally expensive due to multiple light bounces between plies. Thus, we desire an even  simpler model; we aim to represent the cloth solely by yarn curves. Building upon this ply-based approach, we pose the question: \emph{Can we completely conceal the hierarchical structure as well as the light transport within the yarns, treating the yarn as a whole aggregate model that embeds all sub-yarn geometries and light transport?} A yarn-based model is notably more challenging than a ply-based model because simplifying the geometry necessitates a more complicated appearance model. It requires modeling a ray that interacts with a yarn, aggregating light interactions between plies as well as fibers, and ensuring both efficiency and accuracy without introducing obvious trade-offs. 

In this paper, we make the following contributions:
\begin{itemize}
\item \emph{Accurate geometric representation with fiber details}
We propose using yarn curves as input geometry and dynamically computing ply and fiber details during rendering (see Section \ref{sec_geom}). This approach allows us to achieve significant efficiency improvements while maintaining fiber-level details, outperforming previous models such as the ply-based model and fiber-based model \cite{khungurn2015matching}.

\item \emph{An aggregated yarn-based appearance representation}
We introduce a \emph{bidirectional yarn scattering distribution function} (BYSDF), a shading model that computes the appearance of the yarn as a whole using four components, following similar notations as that used in the ply-based model but adjusted for the yarn-level representation (see Section \ref{ssec_bsdf}). These components include reflected and transmitted specular components that capture highlights bouncing off the surface or transmitted through the cloth without much scattering, as well as reflected and transmitted body components that approximate scattered light within the yarn medium. Additionally, we propose a novel consideration for self-shadowing to compensate for the lack of sub-yarn geometry (see Section \ref{ssec_shadowing}).

\item \emph{Multi-scale solution}
Our model begins with a near-field solution to accurately depict the appearance of yarns and seamlessly transitions between near-field and far-field rendering using an efficient integration technique based on pixel coverage (see Section \ref{ssec_multi_res}). Inspired by fur and hair models \cite{chiang2015practical, yan2017efficient}, our approach addresses the issue of resolving individual highlights within yarns when viewed from a distance, which traditionally required an inefficient ray sampling process. Our multi-scale rendering significantly reduces variance and enables faster far-view rendering with substantially fewer samples per pixel, while preserving the same level of quality as our near-field model.

\item \emph{Significant speed-up and efficient memory usage}
By employing our aggregated yarn-based appearance model, on-the-fly geometry realization, and efficient integration for multi-scale rendering, we achieve equal quality taking 1/3 to 1/5 of the rendering time and memory usage than the fiber-based model and factor of 2  improvement over the ply-based model. Furthermore, our multi-scale model delivers an additional 20\% speed-up for distant rendering, maintaining equal quality, as shown in Fig. \ref{fig_teaser}. Exact figures are presented later in \tabref{table_performance}. Our code is publicly available to support future work in the supplementary material and at \url{https://github.com/apoorvkhattar/yarn_model_mitsuba3/tree/master}.

\end{itemize}

\section{Related Work}
\label{sec_related}

\subsection{Surface-based Models}
Early cloth rendering used 2D mesh surfaces with BRDFs \cite{adabala2003real, irawan2012specular, sadeghi2013practical} or BTFs \cite{Dana1999} for fast pipelines, but lacked faithful cloth geometry and fiber details essential for close-up and overall appearance at a distance. These methods also overlooked light transmission, crucial for realism. Microfacet BRDF analysis by Ngan et al. \cite{ngan2005experimental} revealed limitations of BRDF-based approaches and specifically examined the anisotropic nature of velvet fibers. In contrast, our method provides fine details for near-field views and accurate light transmission, addressing these shortcomings.


\subsection{Volumetric Models}
Volumetric models in cloth rendering focus on capturing the geometry of individual fibers. Inspired by the pioneering work of Kajiya et al. \cite{kajiya1989rendering}, these models employ micro-imaging techniques like CT scans to obtain precise fiber geometry. The \emph{radiative transport equation} (RTE) is then used to simulate light interactions within the cloth. 

The RTE was extended by Jakob et al. \cite{jakob2010radiative} for anisotropic cloth by introducing direction-dependent scattering and attenuation to simulate light interactions through anisotropic media such as cloth. Further advances were made proposing micro-appearance models \cite{zhao2011building, zhao2012structure, khungurn2015matching} which relied on the anisotropic RTE and CT images for highly accurate rendering, considering fiber-level interactions. However, volumetric models are slower and more memory-intensive than surface-based models, and challenging for non-static cases due to detailed fiber geometry and light interactions and the use of a practical resolution of volumetric grids. A procedural on-the-fly approach addressed the need for pre-storing curves \cite{Luan2017onthefly} as well as fast rendering under restricted lighting \cite{Khungurn2017fast}. However, acquiring data for volumetric representation techniques remains challenging. Our model offers comparable fidelity but is faster, treating the yarn as a whole and analytically aggregating the expensive fiber interactions.

\subsection{Curve-based Models}
Curve-based models depict cloth surfaces using fiber-modeled curves, utilizing \emph{bi-directional curve scattering distribution functions} (BCSDFs), 
originally introduced for hair strands by Marschner et al. \cite{marschner2003light}. This approach has gained popularity for simulating the appearance of fur, hair, and cloth. Subsequent research has improved upon it by considering scattering events \cite{yan2015physically, yan2017efficient, zhu2022practical}.

An alternative model, proposed by Irawan et al. \cite{irawan2012specular}, is a yarn-based model, leveraging geometric information within yarns to represent cloth appearance. Jin et al. \cite{jin2022woven} propose a differentiable fabric built upon the work of \cite{irawan2012specular} in which  a deep neural network is used to estimate the model parameters with a real photograph as a reference. Fiber-based models \cite{khungurn2015matching} were later introduced to capture detailed cloth appearance considering the scattering of light by individual fibers and later extended to explore the correlation between mechanical simulations and fiber appearance \cite{Montazeri2021mechanics}. However, these models can be inefficient due to the slow construction and traversal of the hierarchy of a large number of fibers and the increased computational complexity of simulating light interactions between fibers. They have been further enhanced to include real-time capabilities through core fiber aggregation \cite{wu2017real} but this technique does not support ray tracing, so it cannot handle realistic lighting simulations. Lastly, a neural-based approach has been recently proposed \cite{soh2023neural} that also fails to capture the fiber details in close-up views.

To strike a balance, Montazeri et al. \cite{montazeri2020practical} proposed a ply-based model for woven cloth which was later extended to knitted cloth \cite{montazeri2021practical}. In this model, individual plies are represented as curves and 1D textures are used to add fiber-level details. It successfully achieves a detailed appearance, ensuring energy conservation during scattering events \cite{zhu2022practical} without the need for a large volume hierarchy for fiber curves.

\subsection{Aggregated Models}
We argue that the ply-based model, as discussed in Section \ref{sec_intro} inefficiently represents yarns, due to the use of explicit curves for individual plies. Yarns typically consist of 3 to 12 plies, with 5-ply yarns being the most common \cite{zhao2012structure}. While offering high quality for near-field views, it exhibits high variance in the far-field due to inherent micro-structure details. To address the ply dependency issue, a new yarn model was proposed that offers ply-level details \cite{Zhu2023Hierarchical}. However, apart from the lack of fiber details, this model can be problematic in both geometry and shading: wrong ply geometry can be produced, especially from a grazing angle; sharp silhouettes of plies and significant color mismatches may appear due to incorrect transparency in  dual-scattering style shading \cite{zinke2008dual}. In contrast, our multi-scale yarn-based model offers a comprehensive solution, effectively managing varied ply counts with fiber details without impacting performance and accuracy across a range of viewing distances (see Section \ref{sec_results}).

\section{Modeling Yarn Geometry}
\label{sec_geom}
\begin{figure}[t!]
    \centering
    \includegraphics[width=\linewidth]{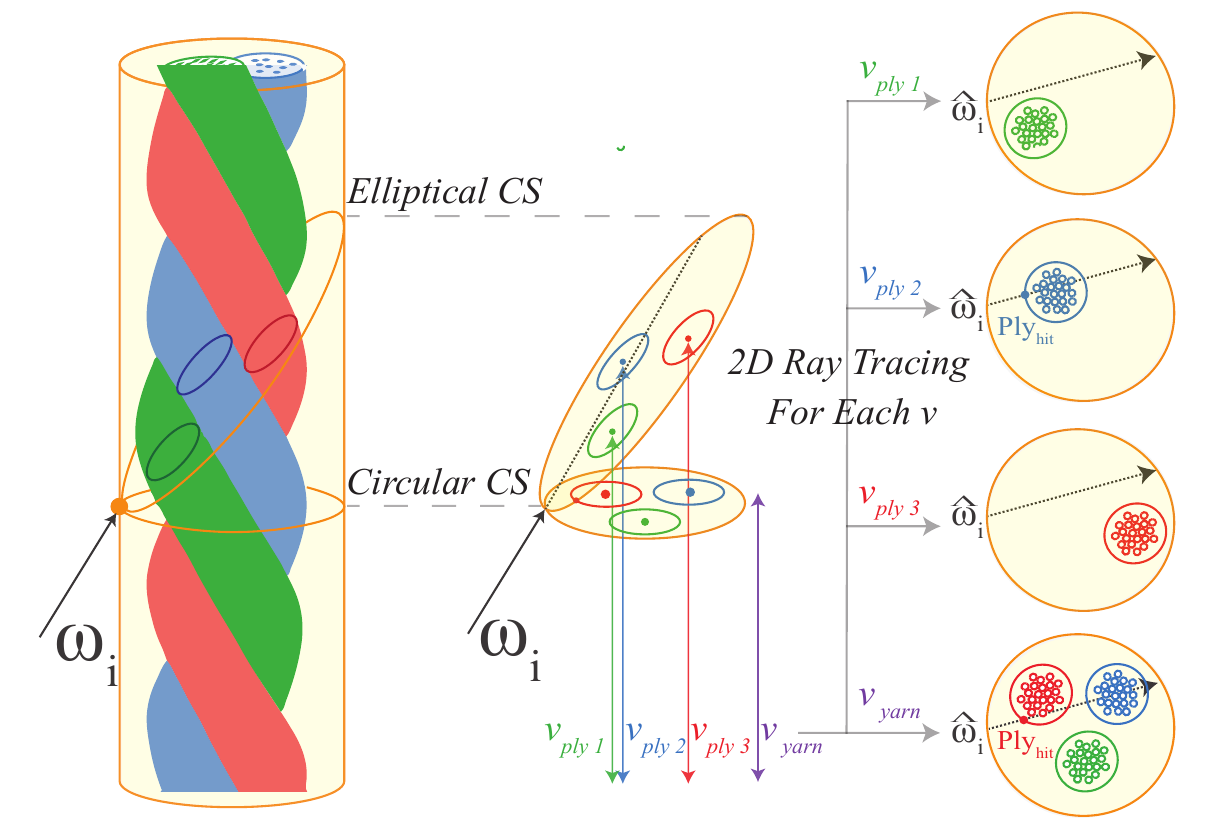}
    \caption{{Implicit tracing using an elliptical yarn cross-section (CS) instead of a circular CS.} The ray first intersects the yarn surface and cuts it into an ellipse. Our iterative approach (discussed in Section \ref{ssec_geom_plies}) calculates the intersection of ply-helices with the elliptical plane and then returns the closest ply by employing 2D ray tracing with a different $\longitudinalLength_\mathrm{ply}$.
    }
    \label{fig_geom3d}
\end{figure}

\subsection{Preliminaries} While the geometric representation of cloth can be complex, its hierarchical structure allows for faithful representation at different scales. Cloth is made up of long strands of \emph{yarn}, consisting of multiple intertwined \emph{plies} and hundreds of micro-diameter twisted \emph{fibers}.

\subsection{Overview}
\label{ssec_geom_overview}
In our model, we focus on accurately representing cloth by explicitly generating yarn curves while implicitly computing ply and fiber details. This approach achieves an efficient yet accurate depiction, particularly well-suited for near-field viewing. Previous models either lack sub-yarn details \cite{sadeghi2013practical}, employ explicit fiber structures that are computationally intensive \cite{khungurn2015matching}, or rely on explicit ply structures that are inefficient for multi-ply yarns \cite{montazeri2020practical, montazeri2021practical}. In what follows, we describe our geometric representation approach for cloth and the implicit ray tracing technique we employ to accurately compute the ray intersections with the cloth sample.

Our simplified geometric method takes a ray and a curved cylinder representing the yarn geometry as inputs. The outputs include the intersection point with the ply and the fiber's canonical frame (normal and tangent) at that point as if the fiber were  intersected explicitly. In case the ray does not find the intersection with the plies after hitting the yarn, a no-hit case is reported as an output. 

\begin{figure}[!t]
    \centering
    \includegraphics[width=\linewidth]{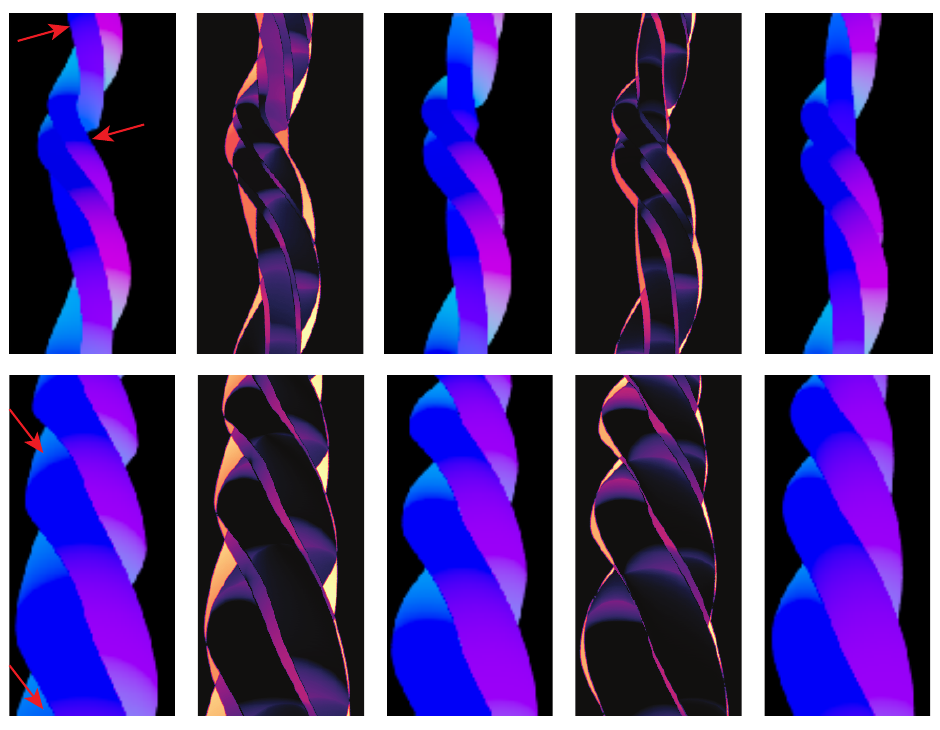}
    \put(-130,-6){\color{black}\small{ Elliptical}}
     \put(-125,-15){\color{black}\small{(Ours)}}
    \put(-84,-6){\color{black}\small{ Elliptical}}
    \put(-86,-15){\color{black}\small{(Error map)}}
    \put(-221,-6){\color{black}\small{ Circular}}
    \put(-176,-6){\color{black}\small{ Circular}}
    \put(-179,-15){\color{black}\small{(Error map)}}
    \put(-34,-6){\color{black}\small{Reference}}  
   
    \caption{{Implicit plies using an elliptical cross-section (CS).} Circular CSs cause stretching or shortening in highly curved yarns (first row), leading to tangents differing significantly from the reference. In a slightly oblique view at 60$^\circ$ (below), the circular approximation introduces inaccuracies, causing an offset in overall shape and tangents compared to the reference, as marked by red arrows.
    }
    \label{fig_elliptical}
\end{figure}

To determine the ply hit point, we propose an implicit ray tracing by focusing on the elliptical cross-section of the yarn instead of traditional circular cross-sections: see \figref{fig_geom3d}. To this end, we assume plies and fibers to be curved cylindrical helices. Thus, a perpendicular cut through the yarn results in a circular cross-section, while an angled cut yields an elliptical cross-section, which consists of conceptual ply and fiber circles. 
By implicitly tracing the ray and analyzing light interactions within the elliptical cross-section formed by the ray, we can extract relevant information about the ply and fibers without the need for complex ray tracing computations involving explicit ply and fiber curves. It is important to note that our novel implicit ray tracing using the elliptical cross-section is more accurate than the traditional circular cross-section perpendicular to the yarn axis \cite{yan2017furbssrdf, Zhu2023Hierarchical} especially when viewed at grazing angles, as demonstrated in \figref{fig_elliptical}.

\subsection{Explicit Yarns}
\label{ssec_geom_yarns}
\begin{figure}[t]
    \centering
    \includegraphics[width=\linewidth]{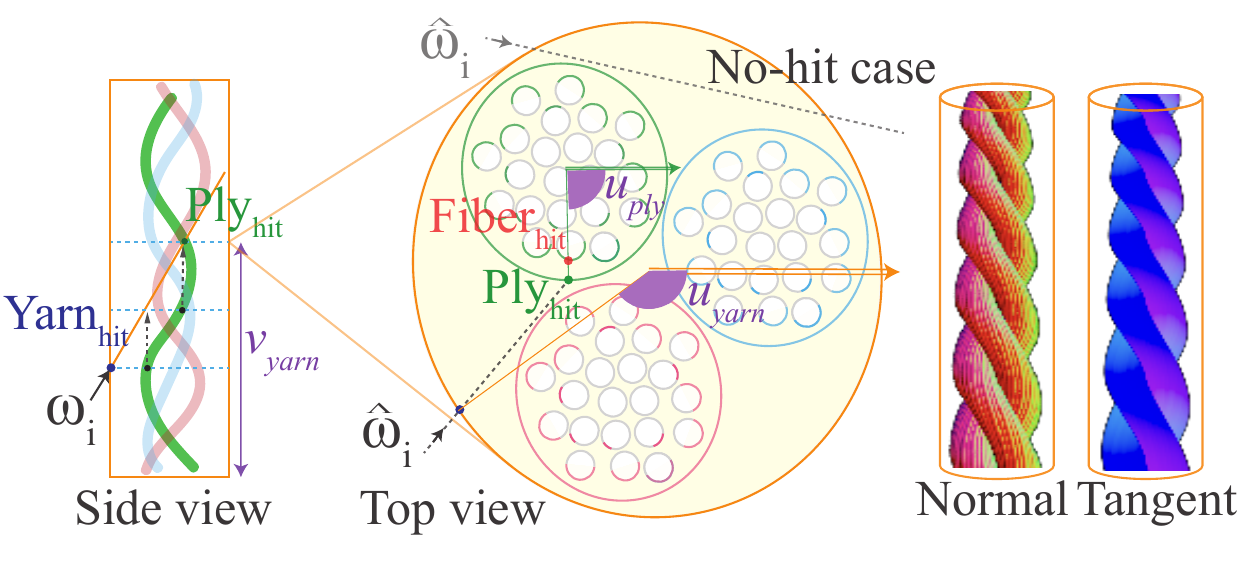}
    \caption{{Schematic overview of our iterative approach to finding the ellipse-helix intersection.} The ray intersects the yarn surface, forming an ellipse in the side view. Newton iteration is used to find the helix-ellipse intersection, determining the ply hit and ply center, obtaining azimuthal phase $\azimuthalPhase_\mathrm{ply}$ and longitudinal length $\longitudinalLength_\mathrm{ply}$ which is used to add fiber texture from 1D texture maps.
    }
    \label{fig_geom}
\end{figure}

We represent the yarns as B-spline curves. The hit point computed for a curved cylinder has a $UV$ map for the yarn, where $\azimuthalPhase_\mathrm{yarn}$ represents the  angle between the yarn normal and a reference yarn normal at the root, referred to as \emph{azimuthal phase}, and $\longitudinalLength_\mathrm{yarn}$ represents the length of the hit point from the root of the curve, referred to as \emph{longitudinal length}. Additionally, we generate reference normals which are  static directions determined at every control point of the yarn by creating rotation minimization frames \cite{wang2008computation}, depicted as a double line in \figref{fig_geom}. This fixed direction along the yarn curve is needed as a reference to ensure the azimuthal angles are consistent through an animation. 

\subsection{Implicit Plies}
\label{ssec_geom_plies}
Given the yarn hit point from Section \ref{ssec_geom_yarns}, we employ a 2D ray tracing technique along the elliptical cross-section of the yarn cylinder to model the ply geometry. Previous work \cite{Zhu2023Hierarchical}  relies on the circular cross-section immediately at the yarn hit to compute the ply intersection. However,  
the distance between this cross-section using the projected incident ray $\widehat\bomi$ and the actual ply intersections with the ellipse plane formed along the incident ray $\bomi$ is not negligible, especially at the grazing angle. Therefore, we utilize the more accurate elliptical cross-section strategy. We visualize a comparison of results using elliptical and circular cross-sections by generating error maps of tangents for implicit plies in \figref{fig_elliptical}. Tangent maps, based on the ply-based model, illustrate the impact on yarn geometry when the incident ray is at an oblique angle, especially for curved yarns shown in  \figref{fig_elliptical}. 

The elliptical approach is challenging because as the ray travels through the yarn volume, the positions of the plies also changes, with each ply following a helix. We use a lightweight Newton iteration method to find the intersection of the ply center line (assumed to be a helix) with the ellipse plane using only 3--5 steps. The process starts with the immediate cross-section at the yarn hit, and we update the cross-section to get closer to the final solution as illustrated in \figref{fig_geom3d}. As in the yarn scenario (see Section \ref{ssec_geom_yarns}), the intersected point on the ply is identified by its azimuthal phase $\azimuthalPhase_\mathrm{ply}$ and longitudinal length $\longitudinalLength_\mathrm{ply}$. This is done for each ply separately until we find the intersection of the ply helices with the ellipse plane.

In tracing for each step, we propose a heuristic: projecting the incident ray $\widehat\bomi$ and finding the ply center at that specific circular cross-section. Then we intersect the computed ply center with the ellipse plane using a line following the yarn tangent direction. At the new position with a different $\longitudinalLength_\mathrm{yarn}$ we repeat the circular processing in the updated cross-section to find the ply center closer to the final solution. Once the ply center converges and there is no change in two consecutive iterations, the final solution of the helix and ellipse is returned as the hit-ply. This offers  fast tracing; please note the  circular cross-section at convergence differs from the perpendicular circular cross-section at the yarn hit.

Once the appropriate positions of the ply centers are identified, we determine the ply closest to the yarn hit point. Having done so, we update its normal ($n_p = n(\azimuthalPhase_\mathrm{yarn})$) and tangent ($t_p = t(\azimuthalPhase_\mathrm{yarn})$) to incorporate the necessary ply-level geometry. At the hit point, the local ply tangent and ply normal are computed to form the ply geometry. The tangent vector is calculated using the first derivative of the 3D helix  at the hit point, and the local ply normal is the vector pointing away from the ply center in the circular cross-section. If the ray does not intersect any plies, the ray is allowed to pass through the yarn cylinder.

\subsection{Implicit Fibers}
\label{ssec_geom_fibers}
\subsubsection{Approach}
To model the fiber geometry, we adopt a precomputation-based approach first introduced in the ply-based model. In this method, the ply cross-section is conceptualized as a collection of individual fiber centers, as illustrated in \figref{fig_geom}. Within this cross-section, we precompute fiber-level details, including normals and tangents, for the outermost fibers. These values are represented as a 1D texture map that wraps around the circumference of the ply and acts as a 1D height-map to capture the visible fibers when viewed uniformly from the ply boundary.

While yarns typically consist of a few plies, each ply is usually composed of tens of fibers, resulting in a tightly packed representation. Therefore, the intersection point at the ply level serves as a reasonable approximation to the nearest fiber, and the necessary fiber-level details can be accessed directly from the texture without the need for additional implicit ray tracing. The resulting normal and tangent directions at the intersection points of the visible fibers are stored in a 1D texture that covers the outermost part of the cross-section. These precomputed values can be queried later using $\azimuthalPhase_\mathrm{ply}$. Finally, by combining the values obtained from Section \ref{ssec_geom_plies} and rotating the ply frame further based on the fiber frame, the final normal and tangent in global space can be computed as: $n_f = n'(R(n_p))$ and $t_f = t'(R(t_p))$, respectively. where $n'$ and $t'$ are 1D perturbations from the precomputed textures, and $R$ represents the rotation of the fibers.

\subsubsection{Adding fiber migration} 
In Section \ref{ssec_geom_fibers}, initially, 
\begin{figure}[t]
    \centering
    \includegraphics[width=0.7\linewidth]{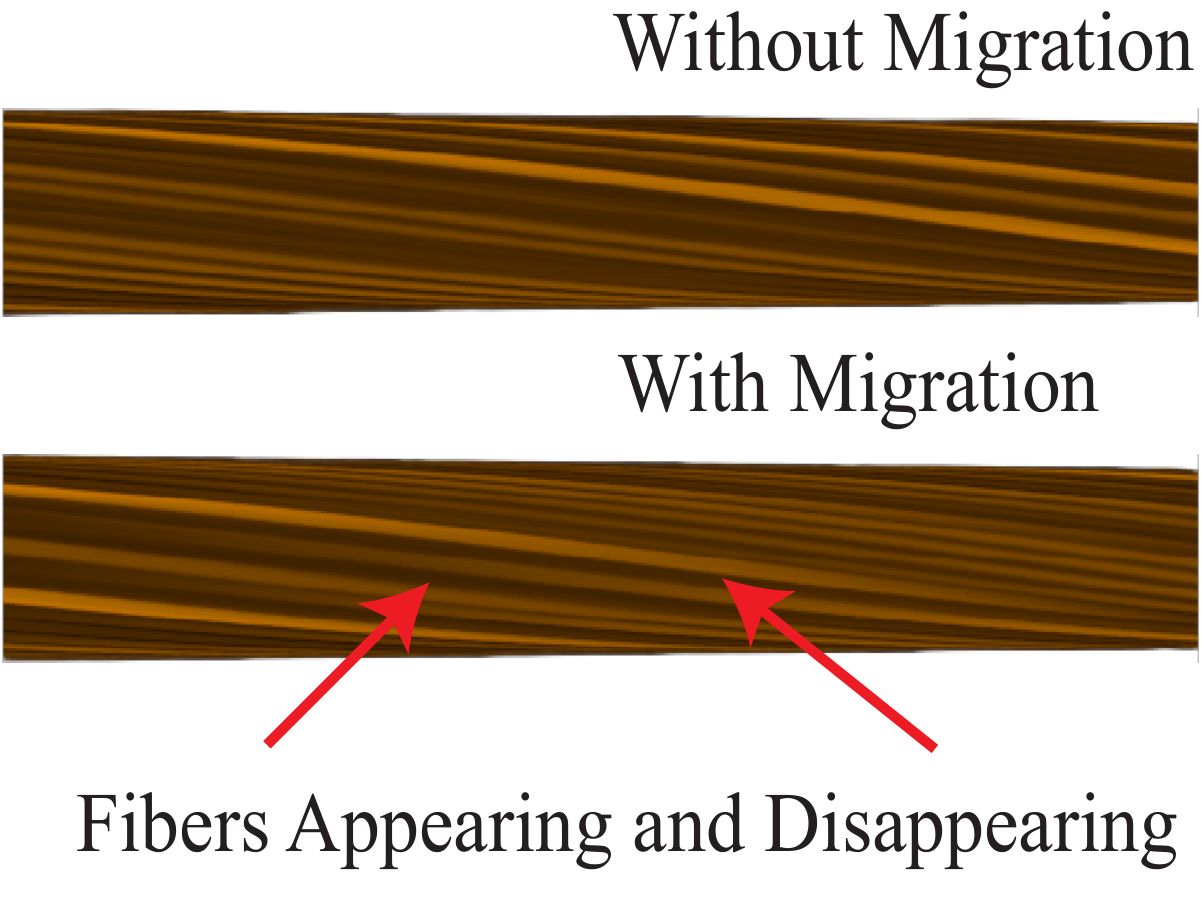}
    \caption{Fiber migration, showing fibers appearing and disappearing to take into irregularities present in a yarn.}\label{rrm1}
\end{figure}
fibers are assumed to follow a regular helical configuration around ply centerlines with a constant radius. However, in reality, fibers exhibit radius variations, known as migration, often characterized by sinusoidal functions \cite{Zhao2016fitting} that sometimes move out and form a loop or disappear as they get closer to the center.   Our implicit ray tracing method, using exact 1D textures from the same fiber distribution in a circular cross-section, introduces an unwanted regular pattern lacking migration irregularities. See Fig.~\ref{rrm1}. To tackle these irregularities, we adopted an approach inspired by the ply-based model. Specifically, we periodically and randomly switch between different 1D textures, formed by distinct fiber alignments in the cross-section. Interpolating between these textures mimics fiber disappearance and introduces irregularities, breaking the continuity.

\section{Modeling Yarn Appearance}
\label{sec_appearance}

In this section, we discuss our approach to modeling the appearance of yarn, which complements the simplified geometry described in Sec. \ref{sec_geom}. Our yarn-based shading model extends the aggregated model first introduced in the ply-based model to capture the appearance of a bundle of fibers.

\subsection{BYSDF}
\label{ssec_bsdf}
\subsubsection{Approach}
The ply-based model provides a plausible shading model for individual plies, which consists of four lobes to capture specular and body components in both forward and backward directions. To adapt this model to yarns with implicit geometry (see Section \ref{sec_geom}), three components require modification to accommodate interactions between plies. Only the immediate reflection component remains the same as in the ply-based model.

\begin{figure}[t]
    \centering
    \includegraphics[width=\linewidth]{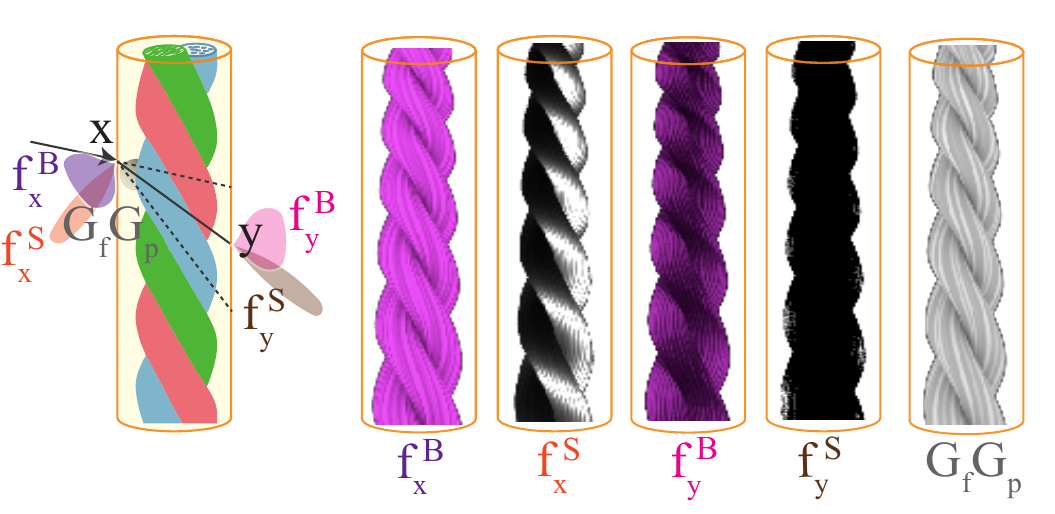}
    \caption{{The components of the BYSDF,}  showing the four lobes and their individual contributions as well as the shadowing component as a product of fiber and ply shadowing queried from the precomputed 1D textures.}
    \label{fig_appearance}
\end{figure}

Following the usual notation from the literature, when an incident ray arrives at the surface of the yarn at point $\bx$, it is divided into forward and backward portions. The backward component captures both the immediate reflection, which is part of the \emph{specular} property $(\bsdfx^\mathrm{S})$, and the scattered light that exits the medium from the same side as the incident ray, referred to as the \emph{body} property $(\bsdfx^\mathrm{B})$. Then, using the transmission component of $\bsdfx^\mathrm{S}$, we sample a point $\by$ on the yarn surface as the exit point, following the GGX distribution. Lastly, the forward component represents both the specularly transmitted light $(\bsdfy^\mathrm{S})$ and the scattered light $(\bsdfy^\mathrm{B})$ that continues in the forward direction. Our yarn-based appearance model is illustrated in Fig. \ref{fig_appearance}.

Assume that we hit the yarn via a path starting from the camera at point $\bx$. $\bsdfx$ is simply the average of two lobes to capture specular and body components but $\bsdfy$ has to collect all contributions from the back side on different $\by$-s because the ray will refract towards different directions into a yarn. Therefore, it can be expressed as an integral with a kernel defined by $\rho$ over $\Omega_{\by}$ which is the part of surfaces that the refracted rays may cover at point ${\by}$. They can be formulated as follows, where $\hat{{\by}}$ is one refracted ray sampled following the GGX distribution: 
\begin{equation}
    \label{eq:fxfy}
    \bsdfx = \bsdfx^\mathrm{S} + \bsdfx^\mathrm{B}, 
    \quad
    \bsdfy = \int_{\Omega_{\by}} \, \left( \bsdf_{\hat{{\by}}}^\mathrm{S} + \bsdf_{\hat{{\by}}}^\mathrm{B} \right) \, \rho_y(\hat{{\by}}) \, d\hat{{\by}}
\end{equation}

\subsubsection{Specular components} 
The specular components $\bsdfx^\mathrm{S}$ and $\bsdfy^\mathrm{S}$ represent the prominent highlights on the fabric surface when light reflects immediately or transmits through the fabric without being scattered. To model these specular lobes, we utilize a rough dielectric BSDF \cite{Heitz2018, walter2007microfacet} with a GGX distribution. The formulations for the specular components at points $\bx$ and $\by$ are as follows, based on $C = \bomi \cdot \bomo$:
\begin{multline}
    \bsdf_{\bx}^\mathrm{S}(\bomi, \bomo) = k_{\bx}^\mathrm{S} \cdot\\
    \begin{cases}
        \dfrac{ {F_{\bx}} \,G_2(\bomi, \bomo, \bomh) \, D(\bomh; \bm{\roughness}_{\bx})}{4 \,|\bomi \cdot \gnorm(\bx)| \, |\bomo \cdot \gnorm(\bx)|} & (C > 0) \\ \\
        \dfrac{|\bomi \cdot \bomt| \, |\bomo \cdot \bomt|}{|\bomi \cdot \gnorm(\bx)| \, |\bomo \cdot \gnorm(\bx)|} &  (C < 0) \\ \\
        \dfrac{\eta^2 \, {(1-F_{\bx})} \, G_2(\bomi, \bomo, \bomt) \, D(\bomt; \bm{\roughness}_{\bx})}{[(\bomi \cdot \bomt) + \eta\,(\bomo \cdot \bomt)]^2}, &
    \end{cases}
    \label{eqn:DxS}
\end{multline} 
\begin{multline}
    \label{eqn:DyS}
    \bsdf_{\by}^\mathrm{S}(\bomi, \bomo) = k_{\by}^\mathrm{S} \cdot \textcolor{gray}{\tau(\bx, \by, \sigT)^{\plyNumIntersected}} \cdot\\
    \begin{cases}
        0, & (C > 0) \\ \\
        \dfrac{|\bomi \cdot \bomt| \, |\bomo \cdot \bomt|}{|\bomi \cdot \gnorm(\by)| \, |\bomo \cdot \gnorm(\by)|} \, & (C < 0) \\ \\
        \dfrac{ {(1-F_{\by})} \, G_2(\bomi, \bomo, \bomt) \, D(\bomt; \bm{\roughness}_{\by})}{[\eta\,(\bomi \cdot \bomt) + (\bomo \cdot \bomt)]^2}, & 
    \end{cases}
\end{multline}
where $\gnorm(\bx)$ is the surface normal function, which is computed using the ply geometry at a given point $\bx$ and similarly for $\by$. The normals are transformed based on the fiber alignments as described in Section \ref{ssec_geom_plies}. We denote this transformation as $\gnorm = \gnorm_f(\gnorm_p)$. $\eta$ represents the refractive index, $F$ is the Fresnel reflection coefficient, $\bomh$ and $\bomt$ are the normalized half vectors for two different cases, and $G_2$ is the Smith uncorrelated masking-shadowing function. The term $D(\cdot; \bm{\roughness})$ represents the normal distribution function (NDF) with $\bm{\roughness} \in \Real^2$ as the (anisotropic) roughness, and $k^\mathrm{S}$ is the specular albedo. 

Notably, a key distinction from the ply-based model is the updated attenuation term $\tau$. The term $\tau(\bx, \by, \sigT)$ considers the attenuation of light between $\bx$ and $\by$ using the Beer-Lambert law \cite{BeerLambert} where $\sigT$ is the material’s extinction coefficient:
\begin{equation}
    \label{eqn:tau}
    \tau(\bx, \by) = \exp(-\sigT \, \| \bx - \by \|).
\end{equation}
In the ply-based model, explicit plies' appearance is studied, and ray tracing multiple scattering events between plies is computationally intensive, particularly for yarns with numerous plies. In contrast, our yarn-based model focuses on modeling the overall yarn appearance, inherently incorporating interactions between plies more efficiently. This is accomplished by computing a new $\tau$ based on $\plyNumIntersected$, the number of intersected plies along the elliptical cross-section for the ray. Thus, as a ray passes through the yarn cylinders, we update the attenuation term as $\tau(\bx, \by, \sigT)^{\plyNumIntersected}$ to account for  attenuation by plies.

\subsubsection{Body components} To capture the scattering behavior of the bundle of fibers as a whole and account for multiple scattering components, we utilize a diffuse-like distribution to approximate the sub-yarn scattering events \cite{yan2017efficient,montazeri2020practical}. At point $\by$, the body component $f_{\by}^\mathrm{B}$ is represented by a Lambertian term. On the other hand, $f_{\bx}^\mathrm{B}$ additionally takes into account a Lommel-Seeliger (LS) term \cite{10.1145/166117.166139, jensen2001practical}, to ensure  energy conservation. The expressions for $f_{\by}^\mathrm{B}$ and $f_{\bx}^\mathrm{B}$ are as follows:
\begin{equation}
\begin{split}
     \bsdf_{\bx}^\mathrm{B}(\bomi, \bomo) = & \, r \, B \, {\plyShadow(\bx, \bomi)} \, \fiberShadow(\bx) \, \\ 
     & k^\mathrm{B}_{\bx} \, \bigg[ \, LS(\bomi, \bomo) \,+ \, \frac{1}{\pi} \bigg],
\end{split}
\label{eqn:DxB}
\end{equation}
\begin{equation}
    \label{eqn:DyB}
    \text{$\bsdf_{\by}^\mathrm{B}(\bomi, \bomo) = (1-r) \, B \, {\plyShadow(\bx, \bomi)} \, \fiberShadow(\bx) \, \frac{k^\mathrm{B}_{\by}}{\pi} $},
\end{equation}
where $k^\mathrm{B}$ is the body albedo and $r$ represents the reflected portion of the body energy $B$, determined in correlation with the refracted albedo of $\bsdf_{\bx}^\mathrm{S}$. The terms $\plyShadow(.,\bomi)$ and $\fiberShadow(.)$ correspond to the ply-shadowing and fiber-shadowing functions, respectively. These functions account for the occlusion and shadowing effects caused by the plies and fibers and are queried from 1D textures as described next.

\subsubsection{Self-Shadowing Components}\label{ssec_shadowing} 
\begin{figure}[ht]
    \centering
    \includegraphics[width=0.7\linewidth]{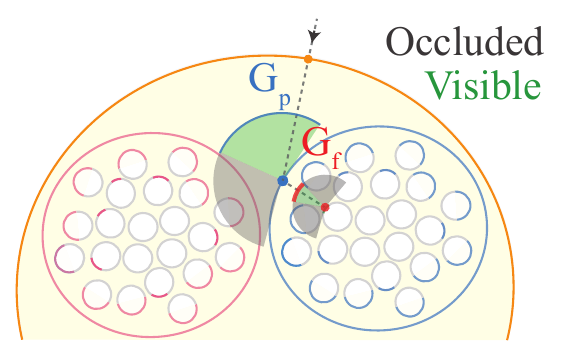}
    \caption{Hierarchical structure for ply and fiber shadows.}
    \label{rrm2}
\end{figure}

To address the limitations of existing yarn-based shading models \cite{irawan2012specular, sadeghi2013practical}, we introduce an additional shadowing component that considers the occlusion caused by one ply on another ($\plyShadow$). See Fig.~\ref{rrm2}.
 In the ply-based method, the ply shadows are handled by the path tracer, thus it only has the fiber shadowing function. In our approach using implicit plies, we approximate  inter-ply shadows since they are part of our yarn appearance. This component is computed by calculating the occlusion ratio at the outermost hit points on the ply surface and storing them in a 1D  texture map. This texture map behaves like a horizon map that wraps around the yarn cross-section circumference. The ply-shadowing module $\plyShadow$ is pre-computed based on the procedural model of the ply structure described in Section \ref{ssec_geom_plies} in the absence of fibers, and it provides full coverage of the yarn surface by sweeping along the yarn centerline.

In addition to the ply-shadowing term, we also consider a fiber-shadowing term $\fiberShadow$ to account for self-shadowing caused by adjacent fibers as elaborated in Section \ref{ssec_geom_fibers}. The multiplication of these two shadowing terms approximates the overall amount of shadowing, compensating for the absence of explicit ply and fiber geometries. Although this shadowing component is computed in a 2D cross-section and does not capture the full 3D occlusion, our experiments have shown that it provides a reasonable estimate of the shadowed regions due to the regular procedural geometric representation of the yarn.


\subsubsection{Importance Sampling}
\label{ssec_importance}
Importance sampling optimizes light interaction in rendering. To manage energy conservation, we define parameter ($B$), to normalize the energy ratio between specularity and body scattering of cloth. This parameter determines specularity energy, with the remaining energy allocated to the body component. The body component is further divided into reflection and transmission using another parameter ($r$), computed based on the probability of the Fresnel term. After selecting the lobe based on energy allocation, importance sampling employs appropriate distributions. The GGX distribution is used for specular lobes, ensuring plausible modeling of their behavior. For diffuse-like components, a cosine-weighted distribution facilitates importance sampling. Details of energy conservation testing can be found in Section \ref{sec_energy}.

\subsection{Adapting to Multiple Scales}
\label{ssec_multi_res}
\begin{figure}[h]
    \centering
    \begin{tabular}{cc}
    \multicolumn{2}{c}{
    \includegraphics[width=0.99\linewidth]{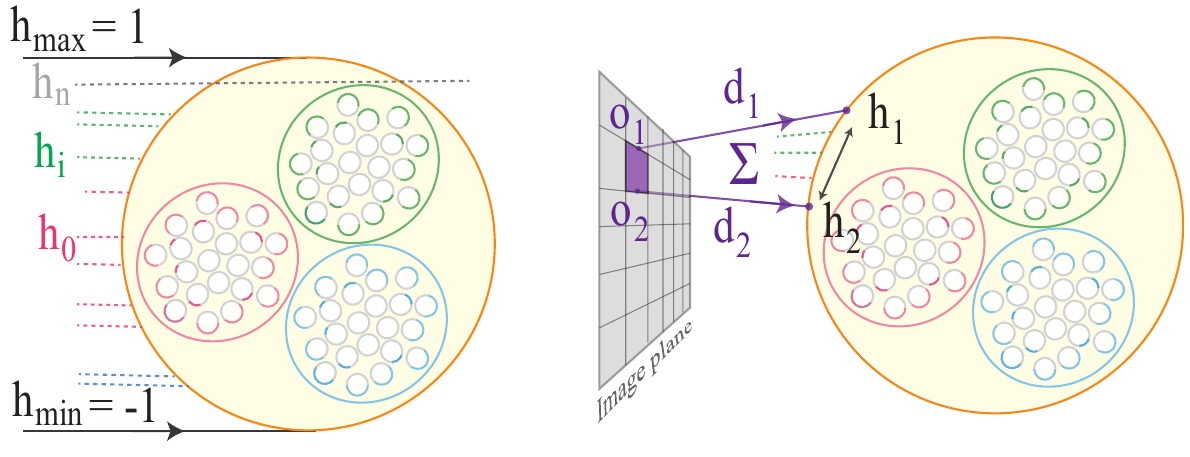}} 
    \end{tabular}
    \caption{ {Adapting to multiple scales.} (a)  far-field integration when the entire azimuthal section of yarn is covered by one pixel. (b)  multi-scale model that uses ray differentials to compute  pixel coverage (purple) on the surface of yarns and to  integrate accordingly to allow a smooth transition from near-field to far-field.}
    \label{fig_multi_scale}
\end{figure}

\begin{figure*}[!t]
    \centering
    \setlength{\tabcolsep}{1pt}
    \begin{tabular}{c ccc ccc}
    & Front lit & Back lit & Front \& back lit &
    Front lit & Back lit & Front \& back lit 
    \\    
     \raisebox{0.03\textwidth}{\rotatebox{90}{Ours}}& 
     \includegraphics[width=0.104\textwidth, angle =90]{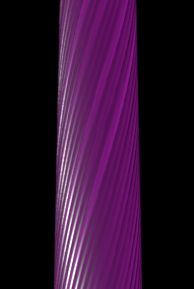} 
    &     
     \includegraphics[width=0.104\textwidth, angle =90]{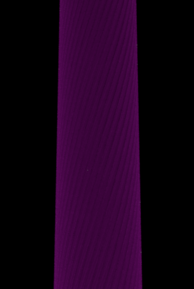} &     
     \includegraphics[width=0.104\textwidth, angle =90]{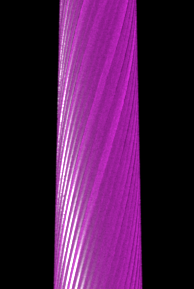} &
     \includegraphics[width=0.104\textwidth, angle =90]{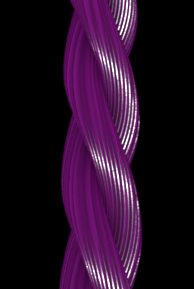}
     &
     \includegraphics[width=0.104\textwidth, angle =90]{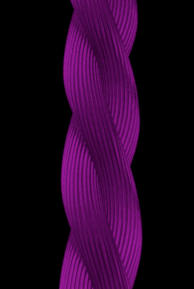} &
     \includegraphics[width=0.104\textwidth, angle =90]{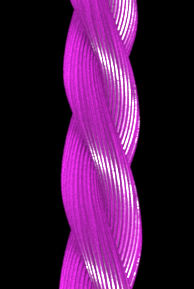}
     \\ 
     \raisebox{0.01\textwidth}{\rotatebox{90}{Reference}} 
     & 
     \includegraphics[width=0.104\textwidth, angle =90]{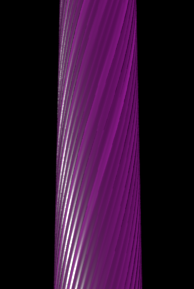} 
     &
     \includegraphics[width=0.104\textwidth, angle =90]{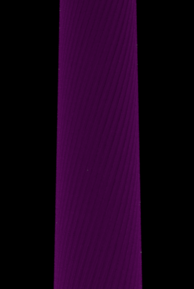} &
     \includegraphics[width=0.104\textwidth, angle =90]{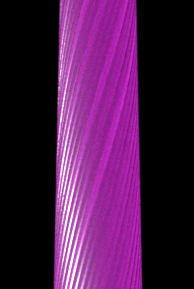} &
     \includegraphics[width=0.104\textwidth, angle =90]{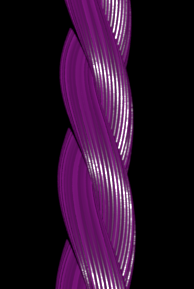} 
     &
     \includegraphics[width=0.104\textwidth, angle =90]{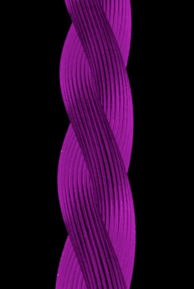} &
     \includegraphics[width=0.104\textwidth, angle =90]{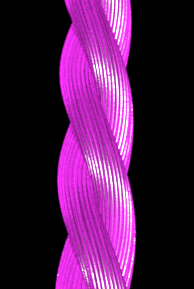}
     \\       
     \raisebox{0.01\textwidth}{\rotatebox{90}{Khung'15}}& 
     \includegraphics[width=0.104\textwidth, angle =90]{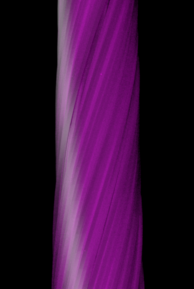}
     &
     \includegraphics[width=0.104\textwidth, angle =90]{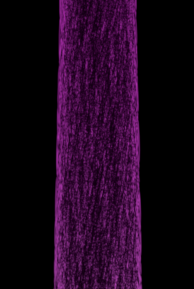}&
     \includegraphics[width=0.104\textwidth, angle =90]{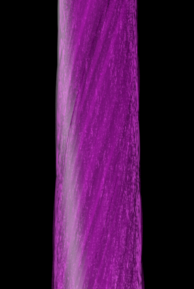}&
     \includegraphics[width=0.104\textwidth, angle =90]{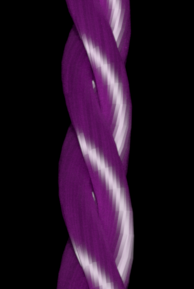}
     &
     \includegraphics[width=0.104\textwidth, angle =90]{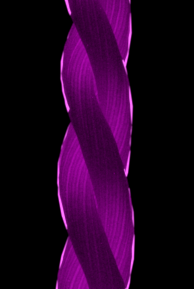}&
     \includegraphics[width=0.104\textwidth, angle =90]{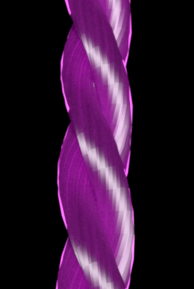}
     \\   
     \raisebox{0.03\textwidth}{\rotatebox{90}{Zhu'23}}& 
     \includegraphics[width=0.104\textwidth, angle =90]{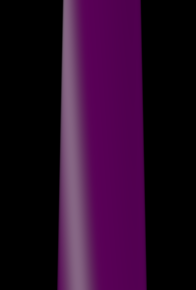}
     &
     \includegraphics[width=0.104\textwidth, angle =90]{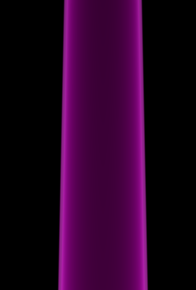}&
     \includegraphics[width=0.104\textwidth, angle =90]{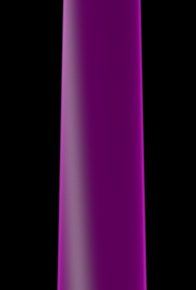}&
     \includegraphics[width=0.104\textwidth, angle =90]{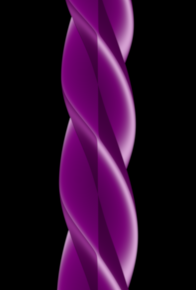}
     &
     \includegraphics[width=0.104\textwidth, angle =90]{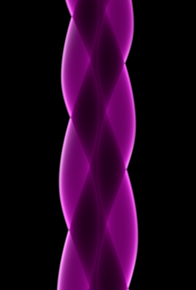}&
     \includegraphics[width=0.104\textwidth, angle =90]{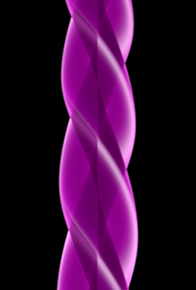}
     \\

     & \multicolumn{3}{c}{-------------- a) Single-ply --------------} 
     & \multicolumn{3}{c}{-------------- b) Three-ply --------------}
     \\
    \end{tabular}
    \caption{\label{fig_micro_comp} {Single yarn comparison} of our approach  to the reference ply-based model, a fiber-based model and a recent yarn-based model for a single yarn to observe the micro-scale events. Quantitative performance is provided in Table \ref{table_performance}.}
\end{figure*}

The  shading model described is efficient for close-up rendering, capturing fiber details in the near field. However, when the camera moves away, the pixel coverage increases, covering multiple fibers or the entire yarn, causing variations in normal distributions and necessitating supersampling to mitigate noise. To address this, we introduce a multi-scale model inspired by Yan et al. \cite{yan2017efficient}, adapting shading computation based on pixel coverage for improved efficiency.

For close-up views covering individual fibers, the original near-field model $\bsdfnear$  in Section \ref{ssec_bsdf} is suitable, considering sub-yarn details. When the camera is distant, and the pixel footage spans the entire yarn, we employ a far-field model $\bsdffar$. In this case, we numerically integrate the BYSDF over the full azimuthal offset range from $\azimuthalOffset_\mathrm{max} = 1$ to $\azimuthalOffset_\mathrm{min} = -1$. This integration captures the overall yarn appearance in the far field, significantly reducing the n umber of ray samples required, compared to the near-field model. By numerically computing the integrals instead of relying on the path-tracer, we achieve more efficient rendering using fewer light samples.

See \figref{fig_multi_scale}. The computation of the far-field model involves stratified Monte Carlo sampling of the range of $\azimuthalOffset$ and integrating accordingly. This approach ensures that the far-field model accounts for the global appearance of the yarn; it can be written as follows with normalization factor $C=\int_{-1}^{1} W(h) \, \mathrm{d}h$, 
\begin{equation}
    \label{eqn:multi_1}
    \bsdffar(\bomi, \bomo) = \frac{1}{C} \, \int_{-1}^{1} W(h) \, \bsdfnear(\bomi, \bomo, \azimuthalOffset) \, \mathrm{d}h.
\end{equation}

To achieve a smooth transition between the near-field model ($\bsdfnear$) and the far-field model ($\bsdffar$), we introduce a multi-scale BYSDF that adapts based on the pixel coverage. The computation of the multi-scale BYSDF relies on ray differentials $\azimuthalOffset_1$ and $\azimuthalOffset_2$. The pixel coverage, represented by the range $(\azimuthalOffset_1, \azimuthalOffset_2) \subseteq (\azimuthalOffset_\mathrm{min}, \azimuthalOffset_\mathrm{max})$, determines the scale of the shading model. When the difference between $\azimuthalOffset_1$ and $\azimuthalOffset_2$ is small, indicating a close alignment, the multi-scale model performs similarly to the near-field model. These ray differentials are explained in detail in  Section \ref{sec_details} and the final formulation of our multi-scale appearance model combines the near-field and far-field components as follows:
\begin{equation}
    \label{eqn:multi_2}
    \bsdfmulti(\bomi, \bomo) = \, \frac{1}{\widehat C} \int_{h_1}^{h_2} W(h) \, \bsdfnear(\bomi, \bomo, \azimuthalOffset) \,  \mathrm{d}h.
\end{equation}
with $\widehat C$ being the normalization over $[h_1,h_2]$. To approximate the multi-scale BYSDF ($\bsdfmulti$), again we use a Monte Carlo algorithm that involves sampling discrete values of $\azimuthalOffset$ within the queried azimuthal range $\azimuthalOffset \in (\azimuthalOffset_1, \azimuthalOffset_2)$. These samples are selected based on their distance to the reference azimuthal offset $\azimuthalOffset_0$ and are used to numerically compute the integrated geometry (Section \ref{sec_geom}) and the shading model (Section \ref{sec_appearance}).

Note that the methods we use to compute $\bsdfmulti$ and $\bsdffar$ are still numerical. However, this approach is much more efficient than relying solely on a ray tracer with only the near-field model. We demonstrate this in \figref{fig_multi_scale}.

To determine the aggregated geometry spanning this range, we compute the weighted average of the normal and tangent vectors of the visible fibers, which are precomputed as 1D texture maps. The weight ($W$) for each $\azimuthalOffset$ represents the distance to $\azimuthalOffset_0$, considering the contribution of the samples based on their distribution. The aggregated normal ($\bm{n}_\mathrm{multi}$) and tangent ($\bm{t}_\mathrm{multi}$) vectors are then used in Eqs. (\ref{eqn:DxS}) and (\ref{eqn:DxB}) to compute $\bsdfmulti$. The computation of $\bm{n}_\mathrm{multi}$ and $\bm{t}_\mathrm{multi}$ can be expressed as follows:
\begin{equation}
    {\bm{n}}_\mathrm{multi} = \frac{1}{\widehat C} \int_{h_1}^{h_2} W(h)  \bm{n}(h) \mathrm{d}h,
\end{equation}
\begin{equation}    
    {\bm{t}}_\mathrm{multi} = \frac{1}{\widehat C} \int_{h_1}^{h_2} W(h)  \bm{t}(h) \mathrm{d}h,
\end{equation}

Similarly, to find the aggregated shadowing component covering the queried azimuthal range, we use the weighted average of the fiber-shadowing terms queried from the precomputed 1D texture map. Aggregated shadowed value $G_\mathrm{multi}$ accounts for the energy loss in the original near-field model due to the presence of the microstructure. Ignoring this term would lead to unwanted brightness, similar to the known issue in microfacet materials. Additionally, the roughness variable of the specular lobe must be also adjusted. Aggregated roughness $\roughness_\mathrm{multi}$ is mapped linearly 
as a function of the azimuthal offset range to compensate for the absence of micro-geometries and the number of samples queried in that range. The updated $\roughness_\mathrm{multi}$ is experimentally given by, $\roughness_\mathrm{multi} = \roughness + (h_2 - h_1)/{\widehat C} $, which naturally contributes to a rougher appearance. 
\begin{equation}
    G_\mathrm{multi} = \frac{1}{\hat C}\int_{h_1}^{h_2} W(h) \, \fiberShadow(h)\, \mathrm{d}h.
\end{equation}
Additional implementation details for the multi-scale feature are explained in Section \ref{ssec_multi_detail}.

\subsection{Energy Conservation Test}
\label{sec_energy}

Our model guarantees energy conservation because the sampling weight to select between the four components of our model as well as each component individually is normalized. In the theoretical case that the albedos are set to 1 and $\sigT$ equals 0, the sum of $f_{\bx}$ and $f_{\by}$ satisfies the criteria for energy conservation, effectively passing the white-furnace test, by construction. When we examine $\bsdf_{\bx}^\mathrm{S} + \bsdf_{\by}^\mathrm{S}$, we observe that it is nearly energy conserving. There is a slight loss of energy due to (i) approximations of multiple scattering within $\bsdf_{\bx}^\mathrm{B}$ and (ii) the absence of internal reflections within $\bsdf_{\by}^\mathrm{S}$, which is set to 0, a limitation also present in previous work. To ensure the entire BYSDF maintains energy conservation, the energy attributed to the missing internal reflection should be equal to $\bsdf_{\bx}^\mathrm{B} + \bsdf_{\by}^\mathrm{B}$. Consequently, there exists a connection between the total scattering coefficient $\sigT$ and the weight of the body components. If $\sigT$ approaches 0, we can expect the body components to tend towards 0 as well, as they primarily result from internal volumetric scattering. 

\begin{figure*}[!t]
    \centering
    \setlength{\tabcolsep}{1pt}
    \begin{tabular}{ccccccc}
    &Front lit & Back lit & Front \& back lit & Front lit & Back lit & Front \& back lit 
    \\    
     \raisebox{0.03\textwidth}{\rotatebox{90}{Ours}}& 
     \includegraphics[width=0.16\textwidth, trim={0cm 4cm 0cm 0cm},clip]{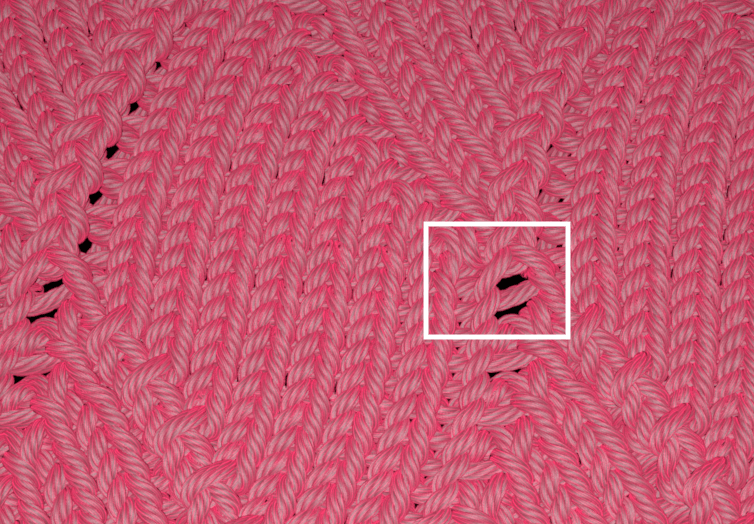} 
     &
     \includegraphics[width=0.16\textwidth, trim={0cm 4cm 0cm 0cm},clip]{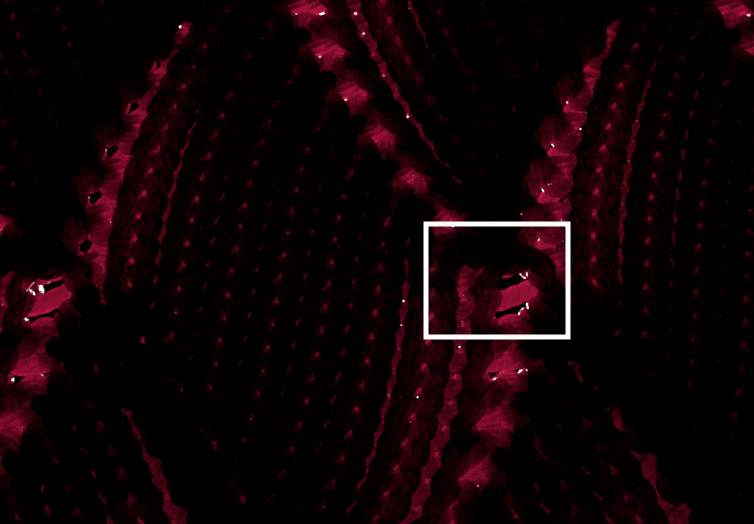} &
     \includegraphics[width=0.16\textwidth, trim={0cm 4cm 0cm 0cm},clip]{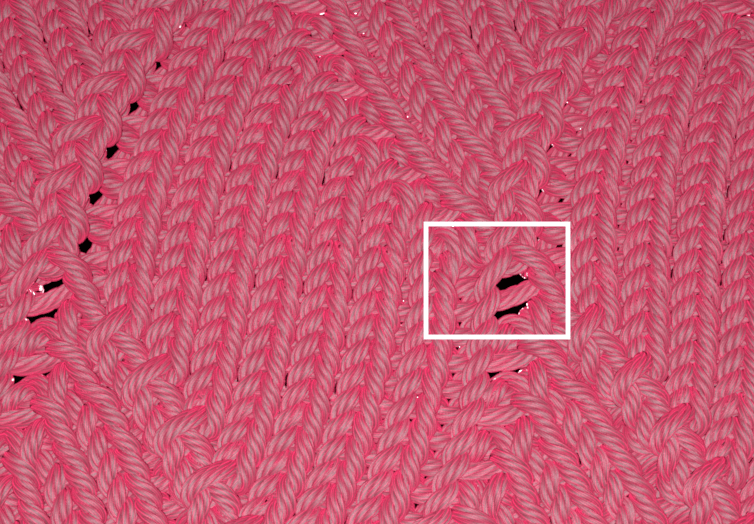} &
     \includegraphics[width=0.16\textwidth, trim={0cm 4cm 0cm 0cm},clip]{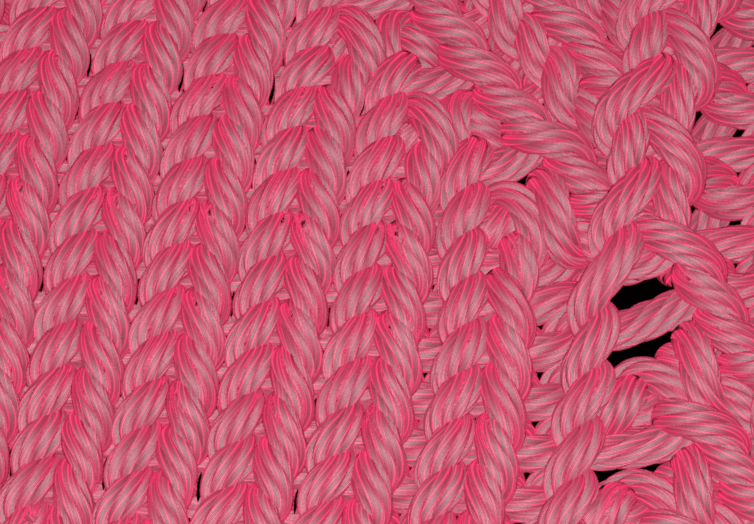} 
     &
     \includegraphics[width=0.16\textwidth, trim={0cm 4cm 0cm 0cm},clip]{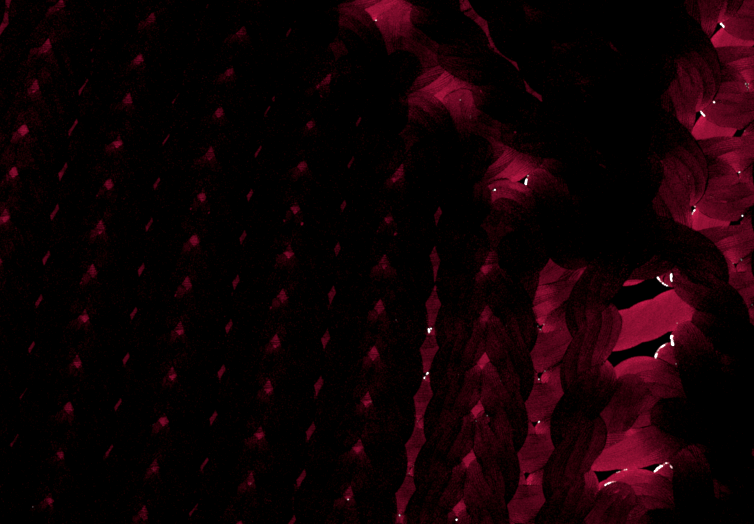} &
     \includegraphics[width=0.16\textwidth, trim={0cm 4cm 0cm 0cm},clip]{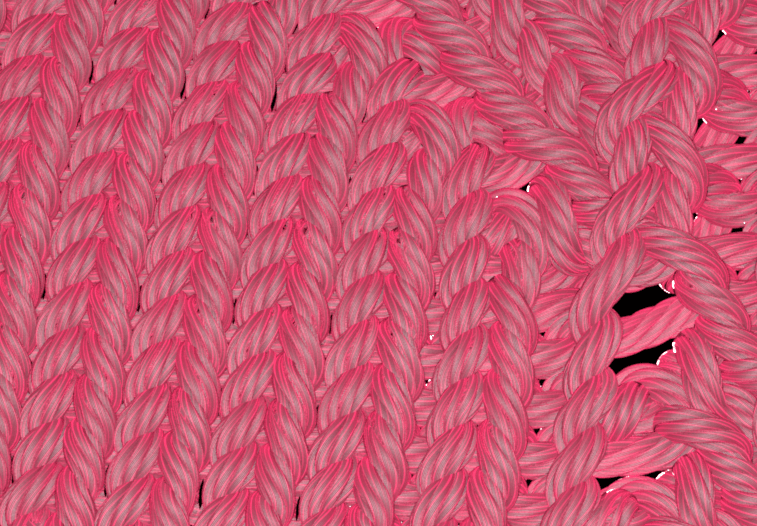} 
     \\ 
     \raisebox{0.01\textwidth}{\rotatebox{90}{Reference}}
     & 
     \includegraphics[width=0.16\textwidth, trim={0cm 4cm 0cm 0cm},clip]{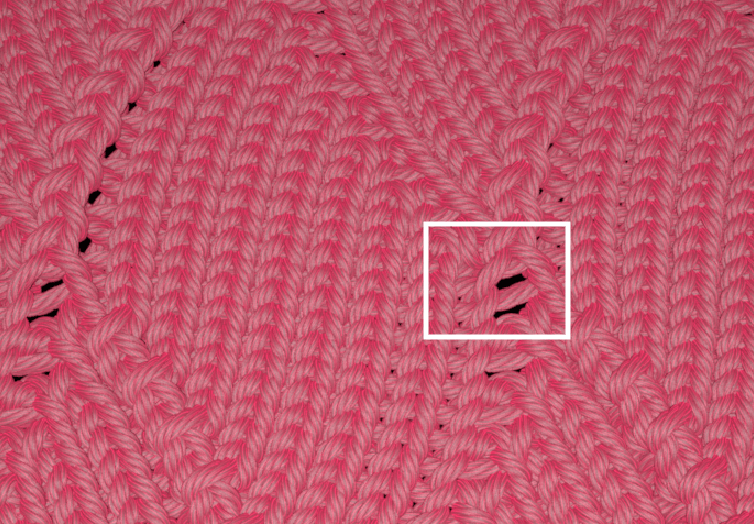} 
     &
     \includegraphics[width=0.16\textwidth, trim={0cm 4cm 0cm 0cm},clip]{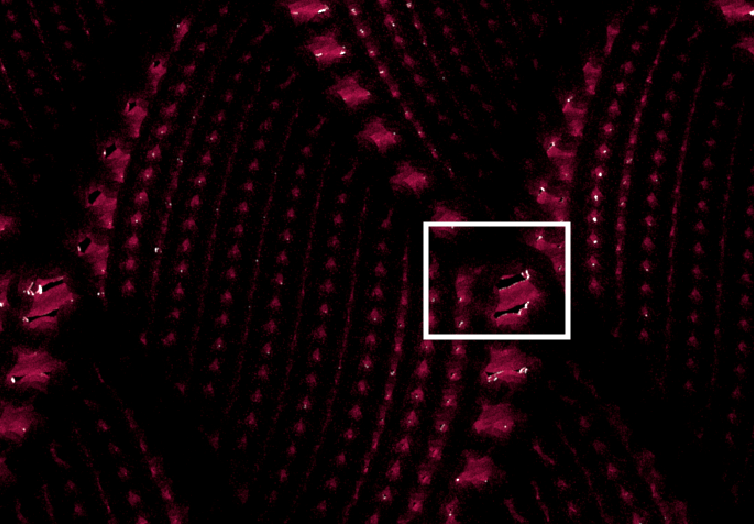} &
     \includegraphics[width=0.16\textwidth, trim={0cm 4cm 0cm 0cm},clip]{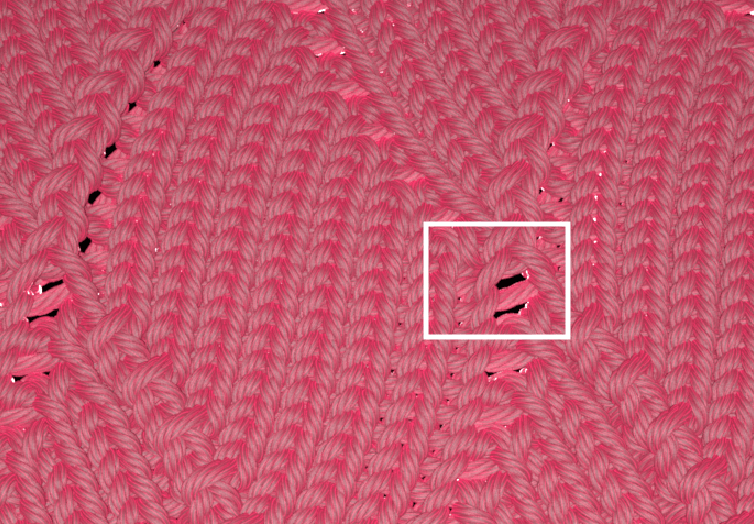} &
     \includegraphics[width=0.16\textwidth, trim={0cm 4cm 0cm 0cm},clip]{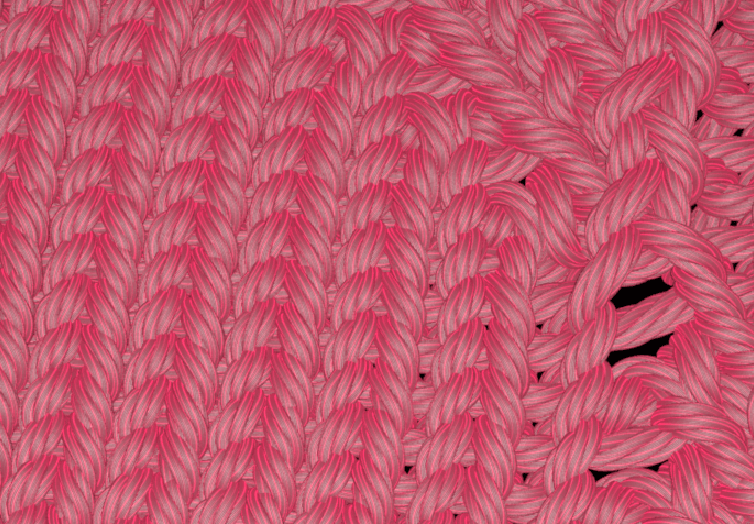} 
     &
     \includegraphics[width=0.16\textwidth, trim={0cm 4cm 0cm 0cm},clip]{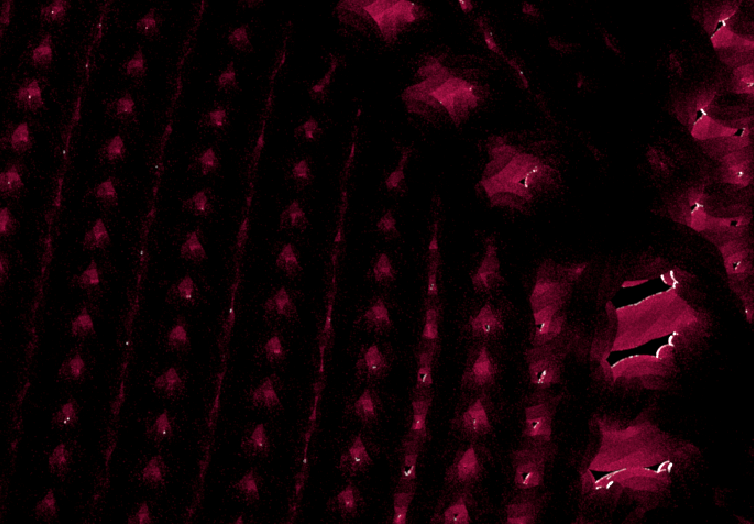} &
     \includegraphics[width=0.16\textwidth, trim={0cm 4cm 0cm 0cm},clip]{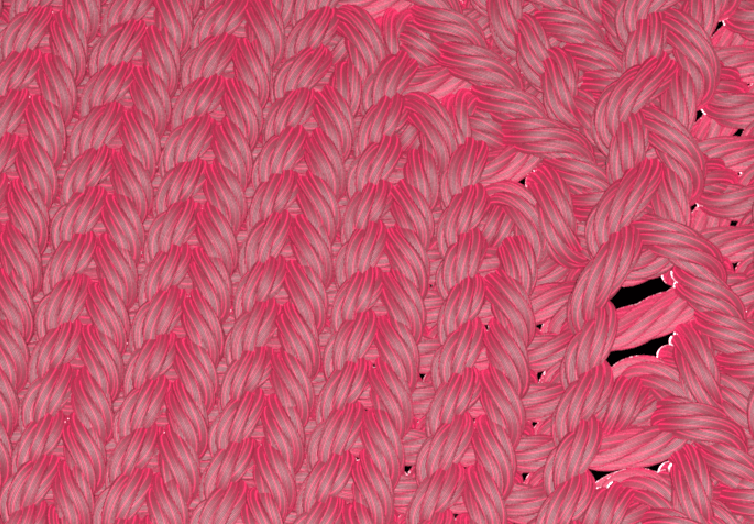} 
     \\ 
     \raisebox{0.02\textwidth}{\rotatebox{90}{Zhu'23}}& 
     \includegraphics[width=0.16\textwidth, trim={0cm 4cm 0cm 0cm},clip]{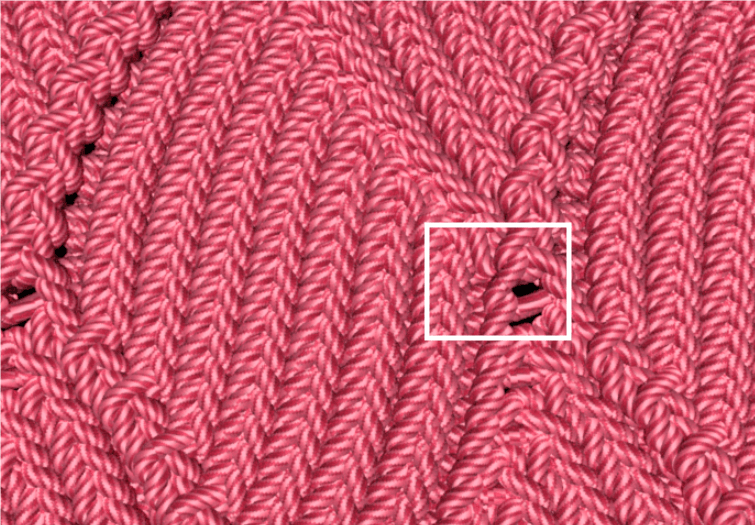} 
     &
     \includegraphics[width=0.16\textwidth, trim={0cm 4cm 0cm 0cm},clip]{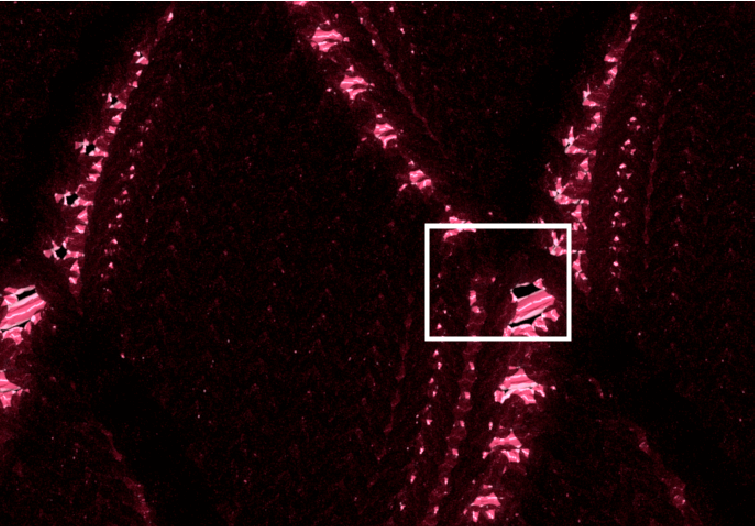}  &
     \includegraphics[width=0.16\textwidth, trim={0cm 4cm 0cm 0cm},clip]{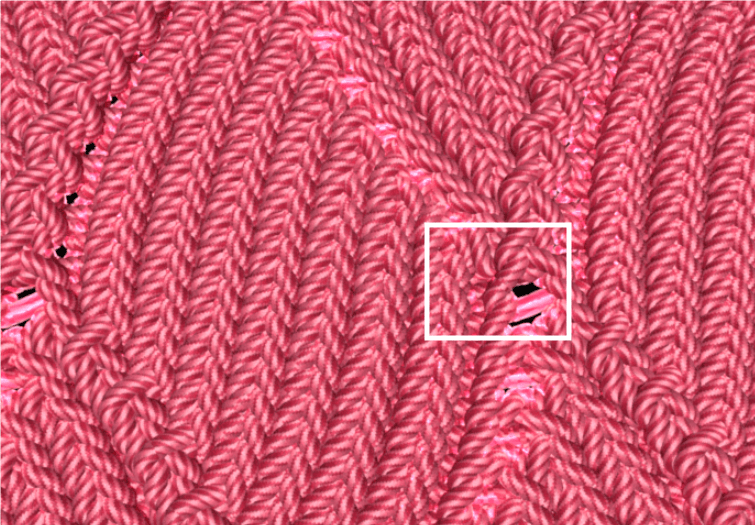}  &
     \includegraphics[width=0.16\textwidth, trim={0cm 4cm 0cm 0cm},clip]{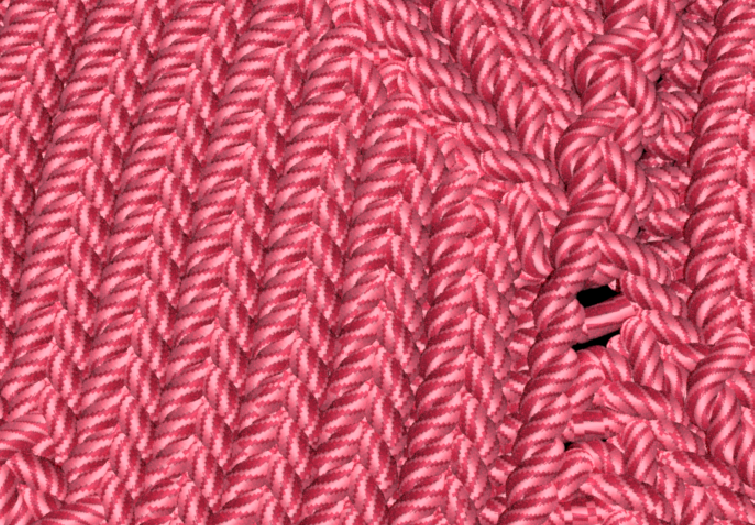} 
     &
     \includegraphics[width=0.16\textwidth, trim={0cm 4cm 0cm 0cm},clip]{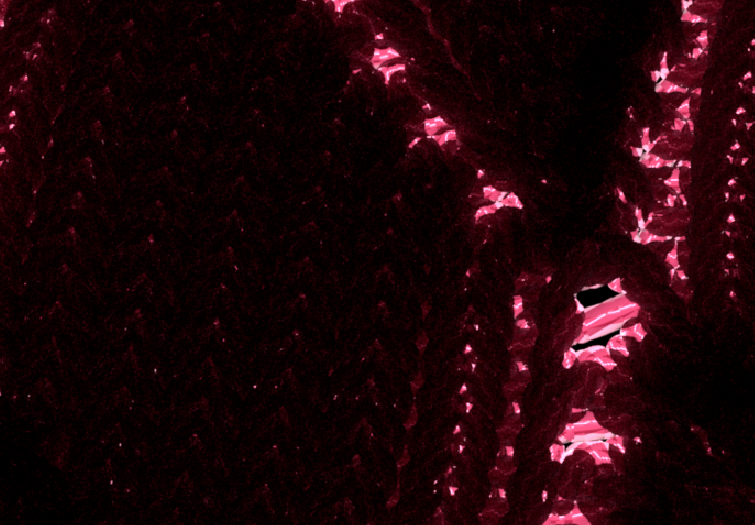}  &
     \includegraphics[width=0.16\textwidth, trim={0cm 4cm 0cm 0cm},clip]{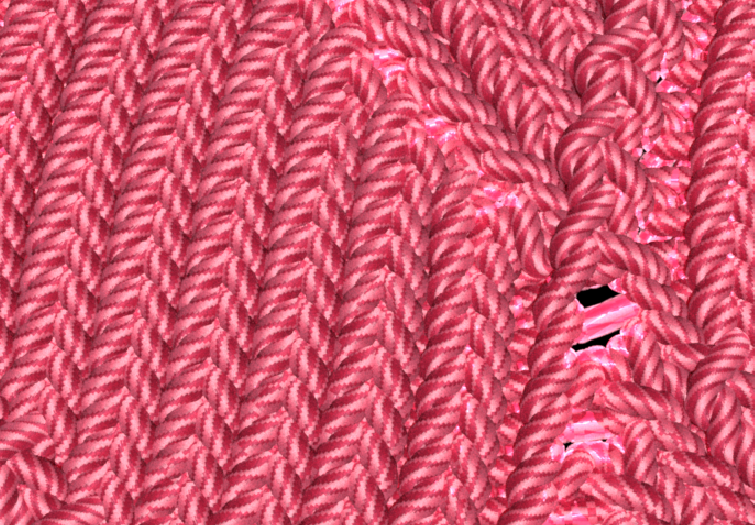} 
     \\ 
     & \multicolumn{3}{c}{-------------- Zoom-out --------------}
     & \multicolumn{3}{c}{-------------- Zoom-in --------------}
     
     \\
     \end{tabular}
    \caption{\label{fig_macro_comp_knit}{Comparison to a hierarchical yarn-based model.} Our model provides a close match to the reference ply-based model, offering faster rendering speed and reduced memory usage for this 6-ply knit sample. Quantitative performance is provided in Table \ref{table_performance}. In comparison to the hierarchical yarn-based model, ours is 1.2 times faster, as well as presenting  fiber-level detail in close-up views. Since the other model cannot accurately handle the geometric information of  multi-ply yarns, the appearance looks more rigid and lacks soft shadows with inaccurate highlights which can be seen in zoomed-in renderings and the focus box in zoomed-out renderings.}
\end{figure*}

\section{Implementation} 
\label{sec_details}
\subsection{Overview}
Our multi-scale BYSDF is implemented as a custom path-tracer in Mitsuba 3 \cite{Mitsuba3}, which supports both environment and local lighting. To efficiently represent the yarn geometry, we utilize the B-spline curve shape module and compute the mapping of  ply and fiber level geometries on-the-fly. 

\subsection{Parameter Fitting}
To provide a realistic appearance, our shading model requires a similar set of parameters   to previous ply-based and fiber-based models. However, our decoupled approach, which separately handles reflection and transmission, provides an advantage in reproducing the target appearance under complex lighting configurations. This decoupling significantly enhances the practicality of BYSDF because the parameters can be fitted more easily. To achieve the best possible match to a target reference and avoid the mundane manual effort of parameter tweaking, we have developed an automatic parameter adjustment method using differentiable rendering. This method allows us to automatically adjust the parameters of our shading model to closely match the appearance of any target reference.

To perform automatic parameter adjustment, we treat the process as an inverse-rendering problem leveraging the differentiable setup in Mitsuba 3, which allows us to compute gradients with respect to the model parameters. Using these gradients, we iteratively adjust and fit the parameters to minimize the pixel-wise mean squared error (MSE) between our rendered image and the target reference. This optimization process ensures that our model captures the desired appearance as closely as possible. We optimize the parameters of the body components i.e. $r$, the diffuse colors $k^\mathrm{B}_{\bx}$ and $k^\mathrm{B}_{\by}$, and the attenuation coefficient $\sigma_t$. Since optimization is sensitive to initial values, the initial values are set the same as those used to generate the ply-based reference. We do not optimize the parameters of the specular component, setting them the same as the reference parameters since they are dependent on the geometry which is handled by the geometric component.

While many parameters can be adjusted independently, implicit relationships exist between parameters like B, $\plyShadow(\bx, \bomi)$, $\fiberShadow(\bx)$, etc. This provides artistic control but can lead to physically implausible choices. Inverse rendering helps find sensible parameters and implicit coupling can constrain optimization to more physically plausible solutions, reducing issues related to ill-posedness and non-uniqueness.

\begin{figure*}[!t]
    \centering
    \setlength{\tabcolsep}{1pt}
    \begin{tabular}{ccccccc}
    &Front lit & Back lit & Front \& back lit & Front lit & Back lit & Front \& back lit
     \\
     \raisebox{0.02\textwidth}{\rotatebox{90}{Ours}}& 
     \includegraphics[width=0.16\textwidth, trim={0cm 4cm 0cm 0cm},clip]{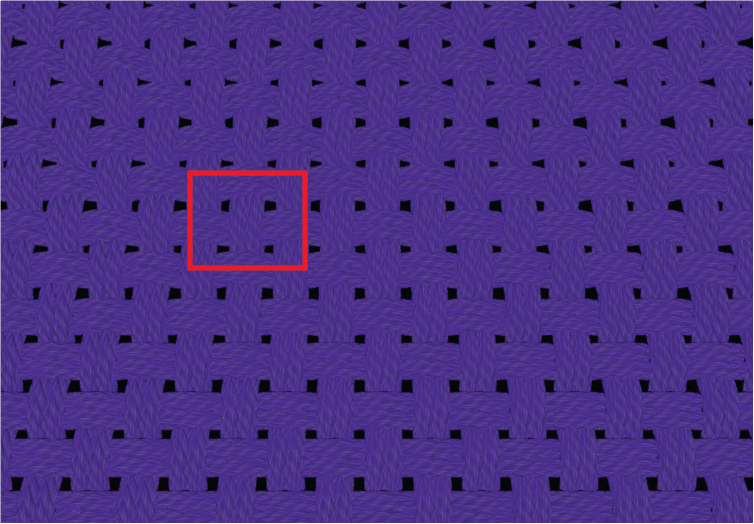} 
     &
     \includegraphics[width=0.16\textwidth, trim={0cm 4cm 0cm 0cm},clip]{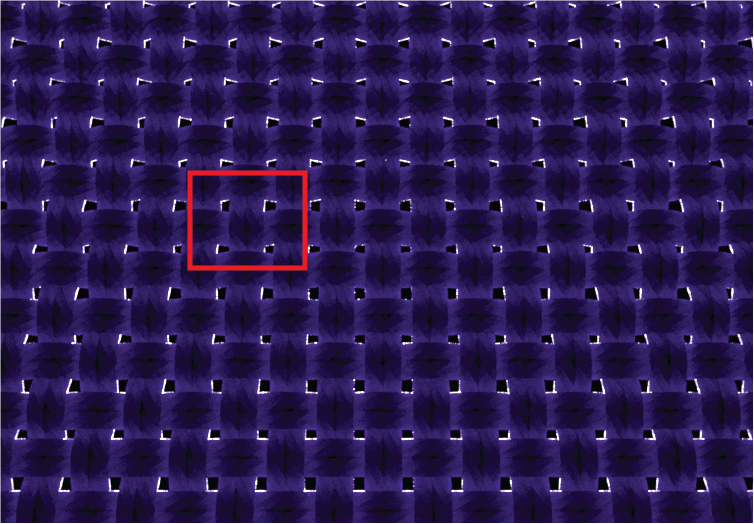} &
     \includegraphics[width=0.16\textwidth, trim={0cm 4cm 0cm 0cm},clip]{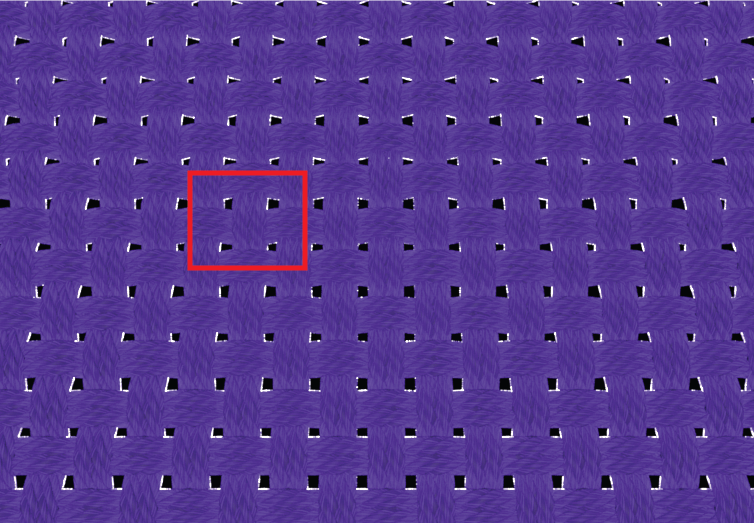} 
     &
     \includegraphics[width=0.16\textwidth, trim={0cm 4cm 0cm 0cm},clip]{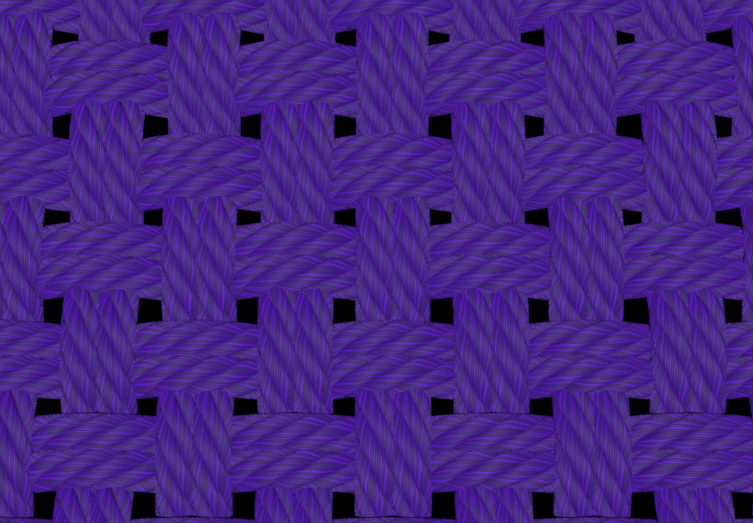} 
     &
     \includegraphics[width=0.16\textwidth, trim={0cm 4cm 0cm 0cm},clip]{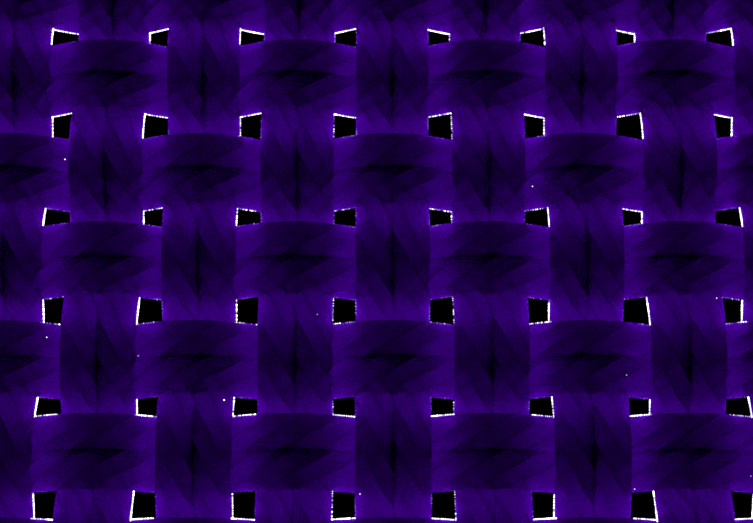} &
     \includegraphics[width=0.16\textwidth, trim={0cm 4cm 0cm 0cm},clip]{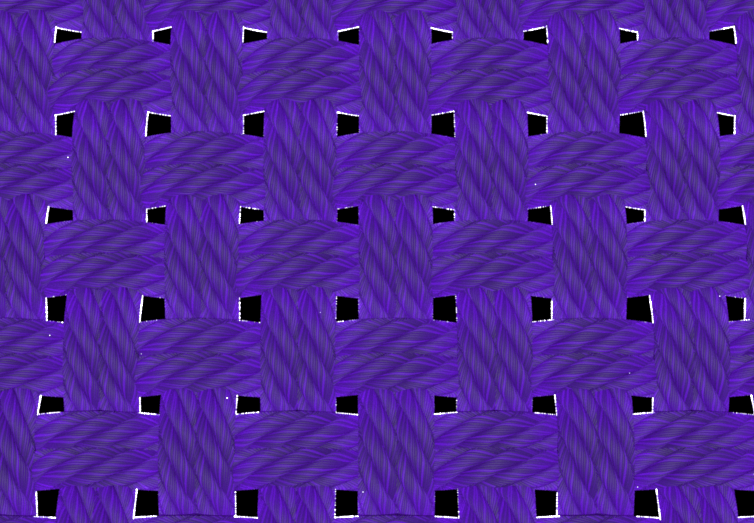}
     \\ 
     \raisebox{0.01\textwidth}{\rotatebox{90}{Reference}}
     & 
     \includegraphics[width=0.16\textwidth, trim={0cm 4cm 0cm 0cm},clip]{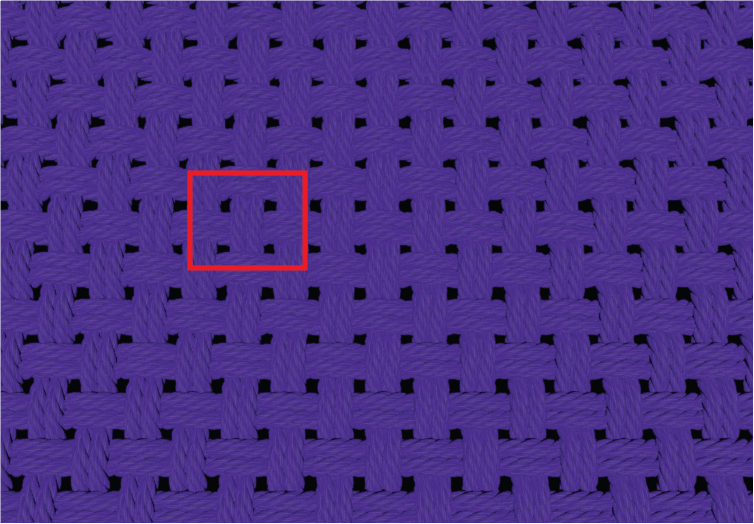}
     &
     \includegraphics[width=0.16\textwidth, trim={0cm 4cm 0cm 0cm},clip]{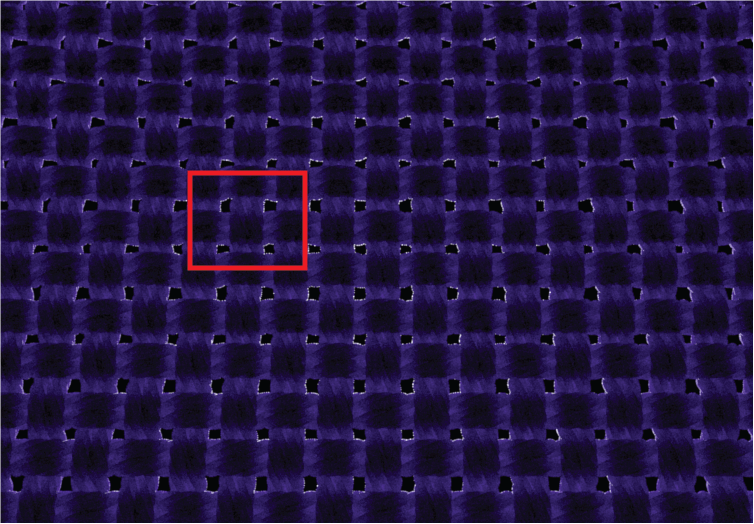} &
     \includegraphics[width=0.16\textwidth, trim={0cm 4cm 0cm 0cm},clip]{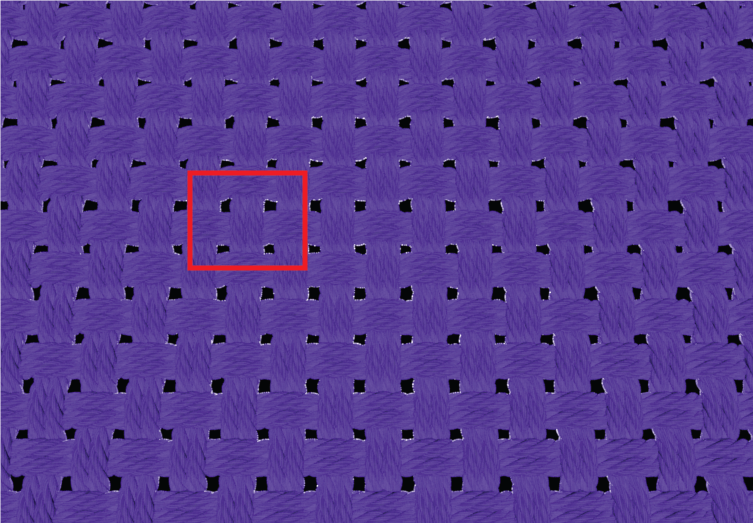} 
     &
     \includegraphics[width=0.16\textwidth, trim={0cm 4cm 0cm 0cm},clip]{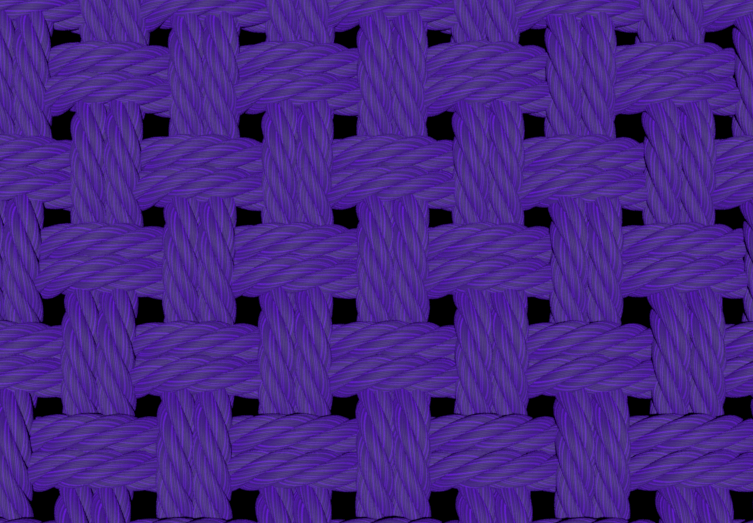} 
     &
     \includegraphics[width=0.16\textwidth, trim={0cm 4cm 0cm 0cm},clip]{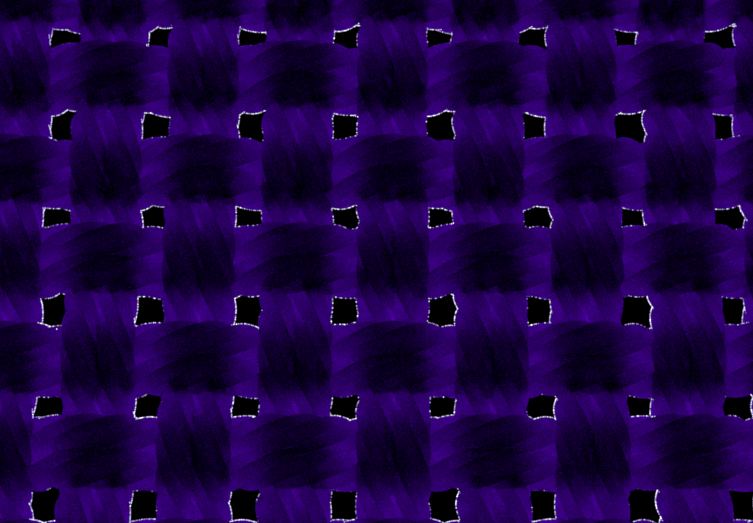} &
     \includegraphics[width=0.16\textwidth, trim={0cm 4cm 0cm 0cm},clip]{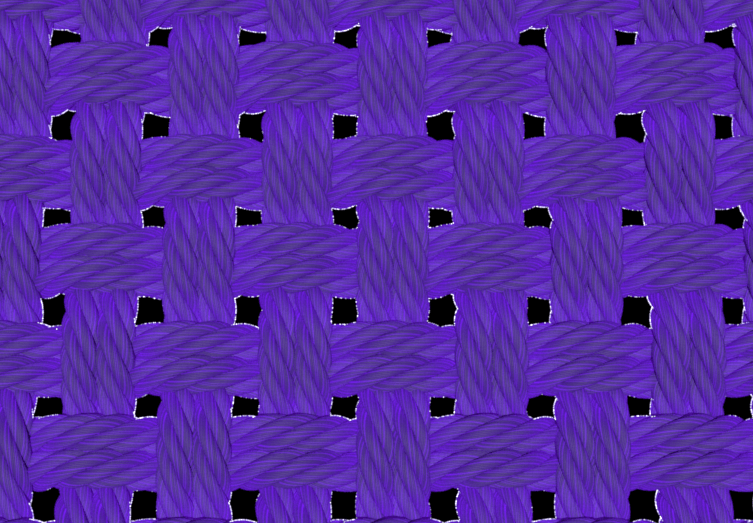} 
     \\ 
     \raisebox{0.0\textwidth}{\rotatebox{90}{Khung'15}}& 
     \includegraphics[width=0.16\textwidth, trim={0cm 4cm 0cm 0cm},clip]{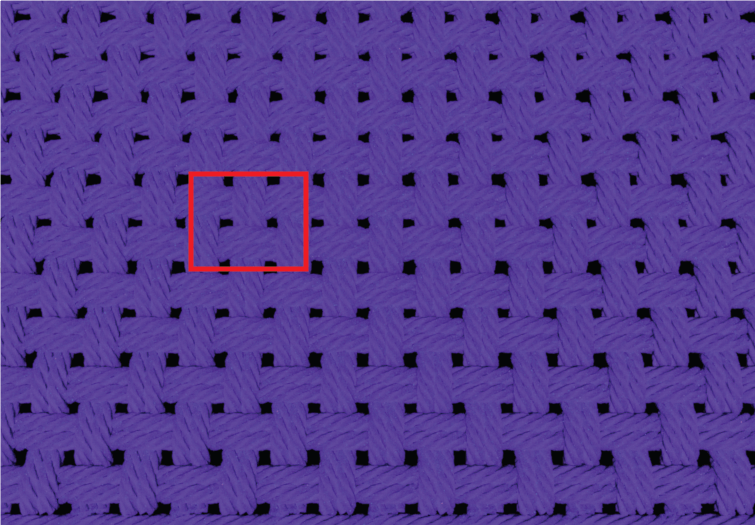} 
     &
     \includegraphics[width=0.16\textwidth, trim={0cm 4cm 0cm 0cm},clip]{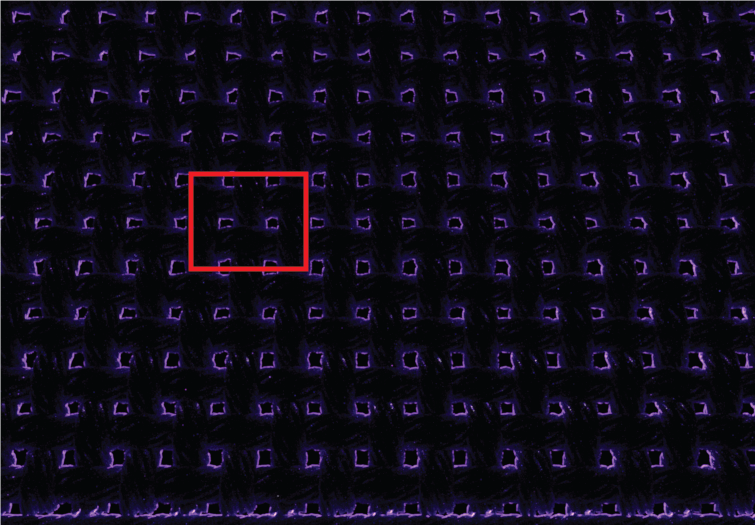} &
     \includegraphics[width=0.16\textwidth, trim={0cm 4cm 0cm 0cm},clip]{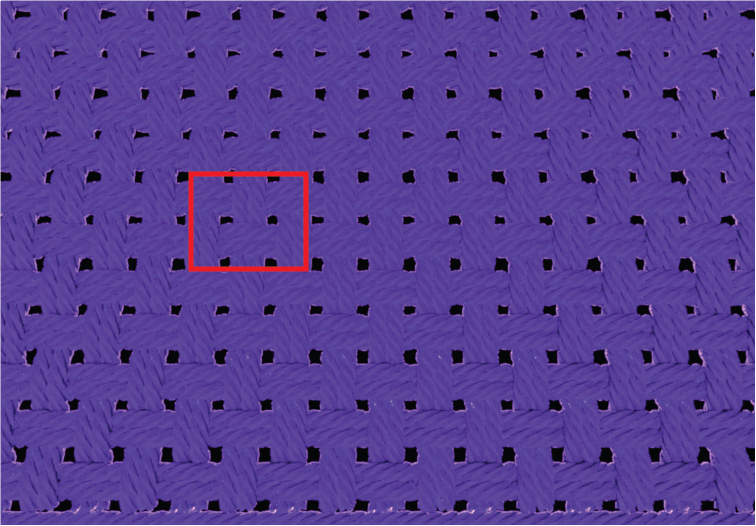} 
     &
     \includegraphics[width=0.16\textwidth, trim={0cm 4cm 0cm 0cm},clip]{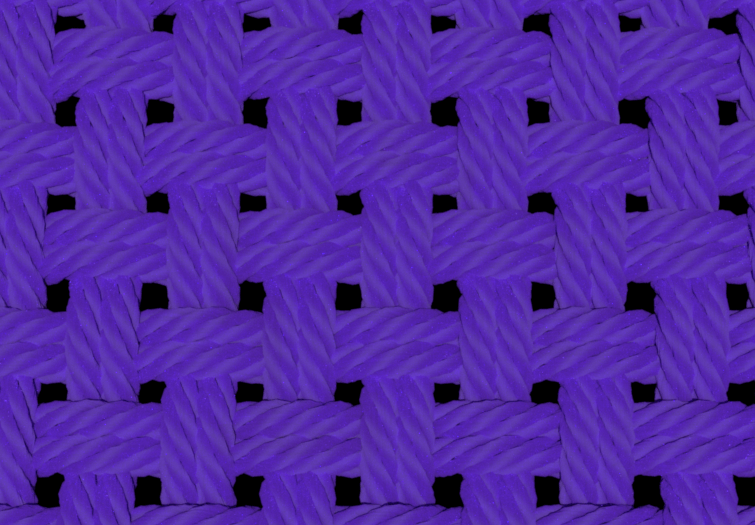} 
     &
     \includegraphics[width=0.16\textwidth, trim={0cm 4cm 0cm 0cm},clip]{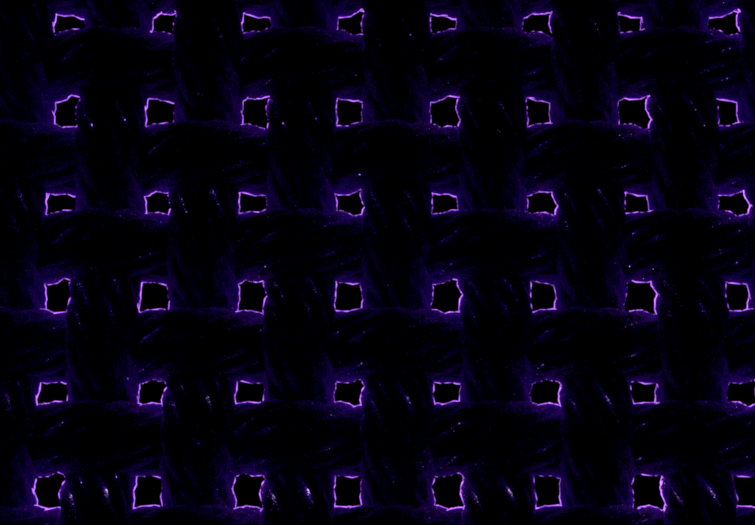} &
     \includegraphics[width=0.16\textwidth, trim={0cm 4cm 0cm 0cm},clip]{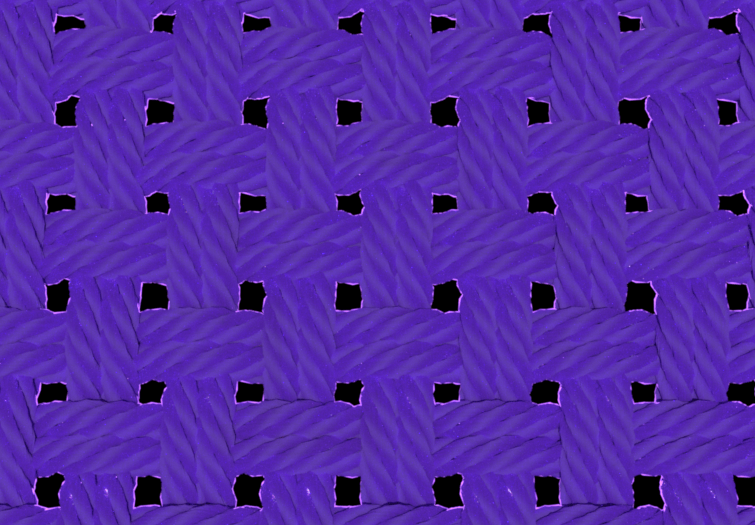} 
     \\ 
     \raisebox{0.02\textwidth}{\rotatebox{90}{Ours}}& 
     \includegraphics[width=0.16\textwidth, trim={0cm 4cm 0cm 0cm},clip]{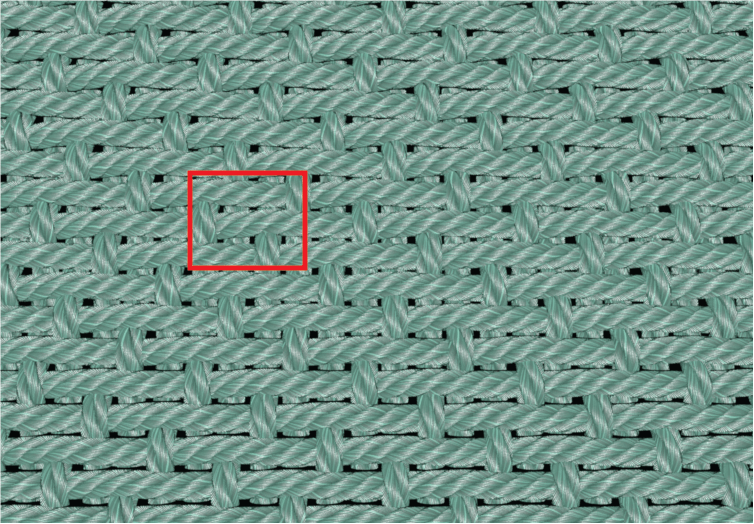} 
     &
     \includegraphics[width=0.16\textwidth, trim={0cm 4cm 0cm 0cm},clip]{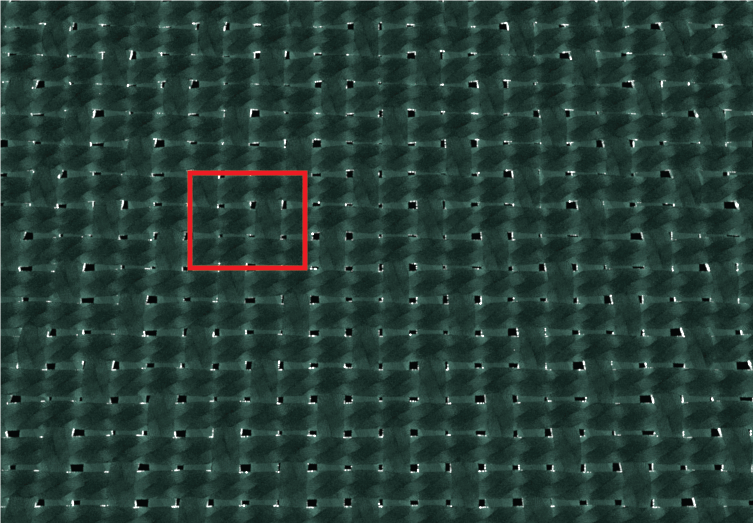} &
     \includegraphics[width=0.16\textwidth, trim={0cm 4cm 0cm 0cm},clip]{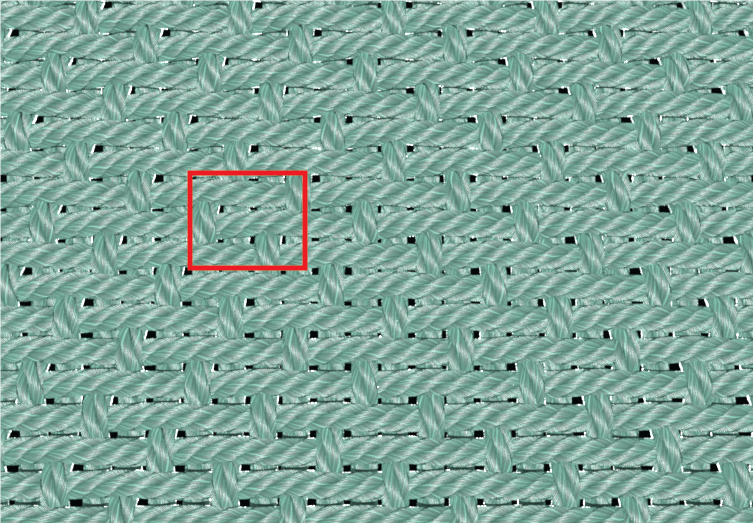} &
     \includegraphics[width=0.16\textwidth, trim={0cm 4cm 0cm 0cm},clip]{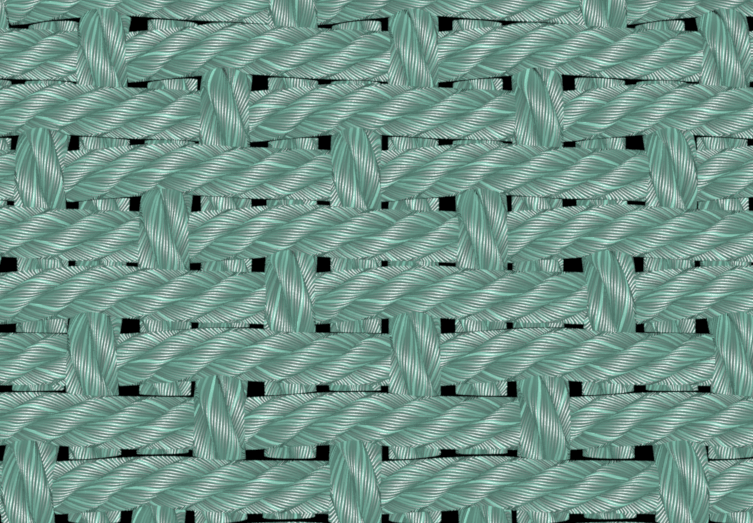} 
     &
     \includegraphics[width=0.16\textwidth, trim={0cm 4cm 0cm 0cm},clip]{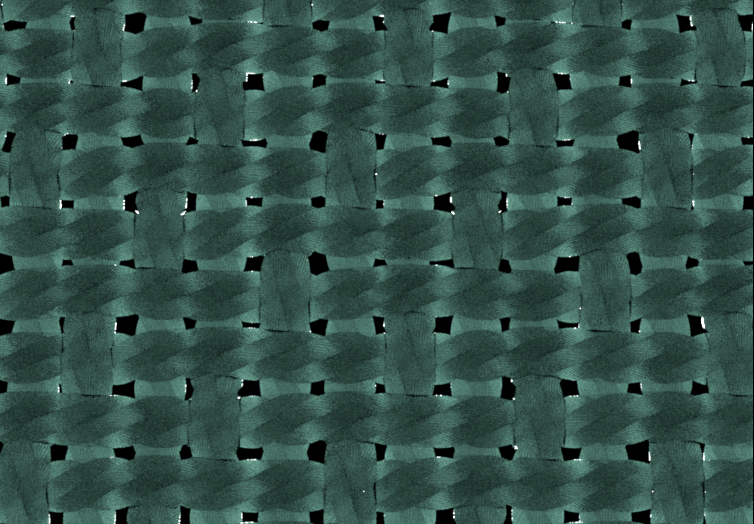} &
     \includegraphics[width=0.16\textwidth, trim={0cm 4cm 0cm 0cm},clip]{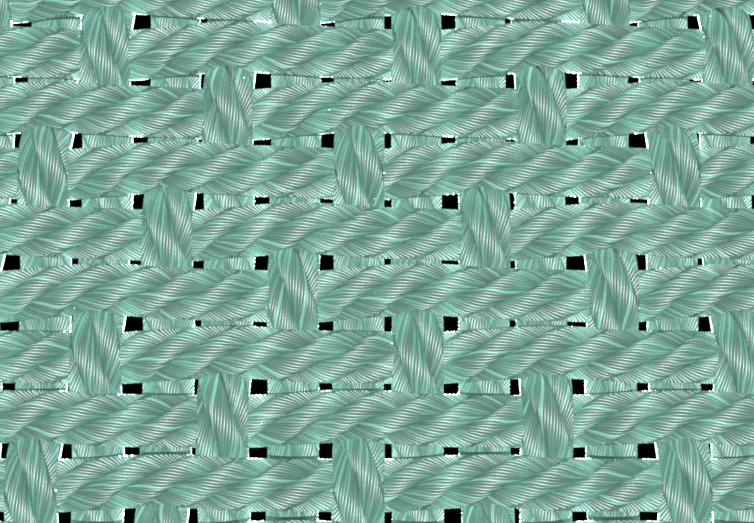}
     \\ 
     \raisebox{0.01\textwidth}{\rotatebox{90}{Reference}}& 
     \includegraphics[width=0.16\textwidth, trim={0cm 4cm 0cm 0cm},clip]{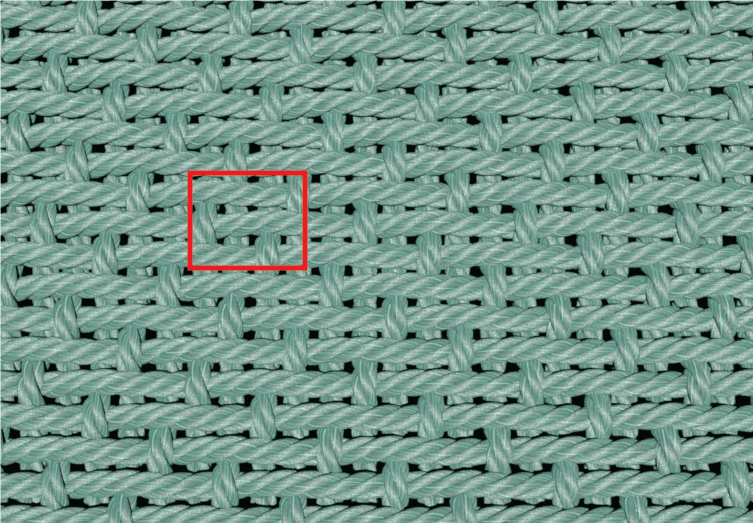} 
     &
     \includegraphics[width=0.16\textwidth, trim={0cm 4cm 0cm 0cm},clip]{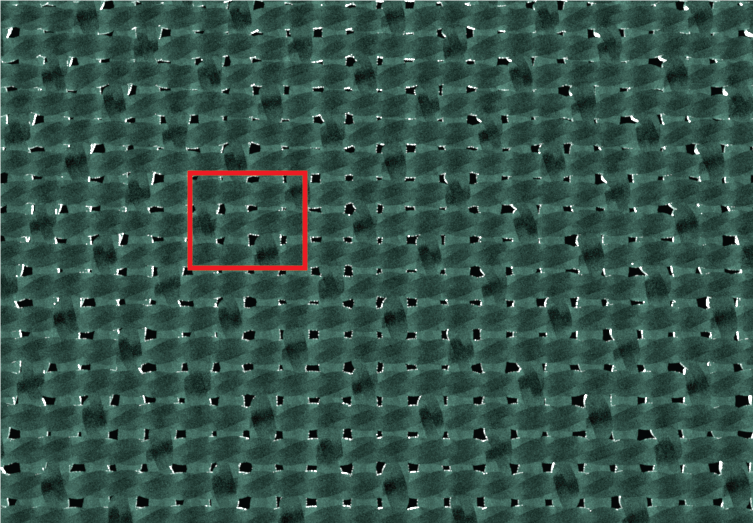} &
     \includegraphics[width=0.16\textwidth, trim={0cm 4cm 0cm 0cm},clip]{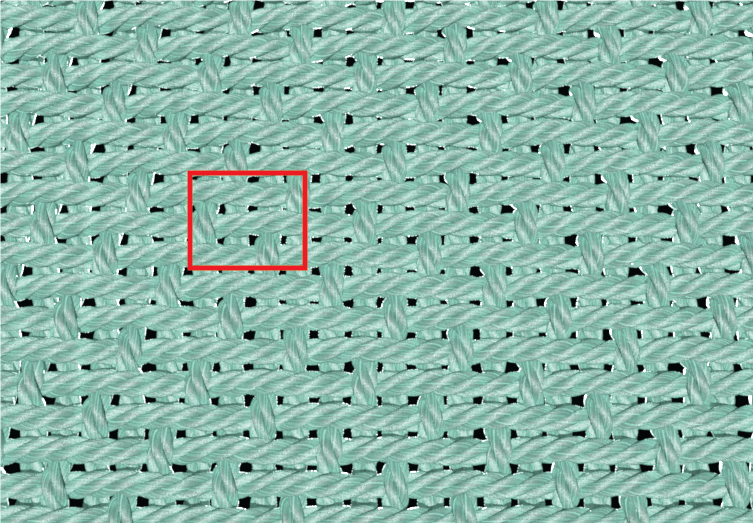} &
     \includegraphics[width=0.16\textwidth, trim={0cm 4cm 0cm 0cm},clip]{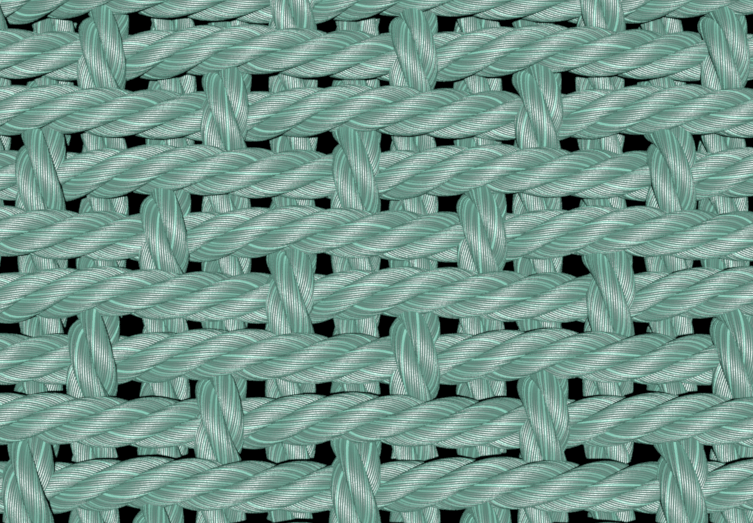} 
     &
     \includegraphics[width=0.16\textwidth, trim={0cm 4cm 0cm 0cm},clip]{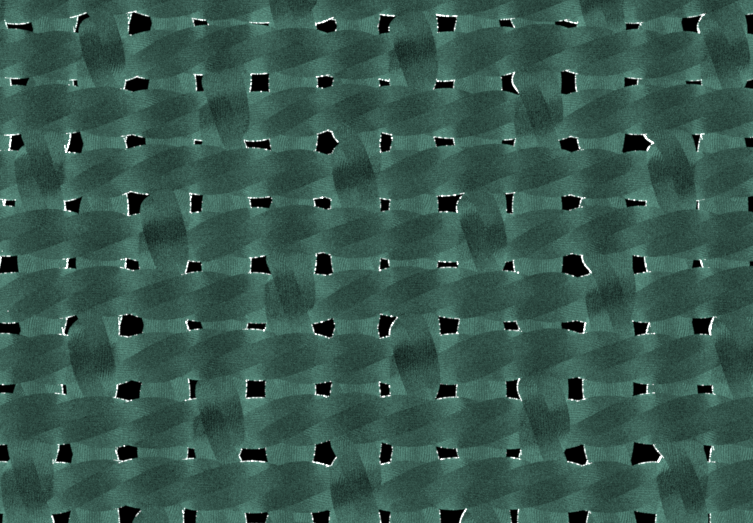} &
     \includegraphics[width=0.16\textwidth, trim={0cm 4cm 0cm 0cm},clip]{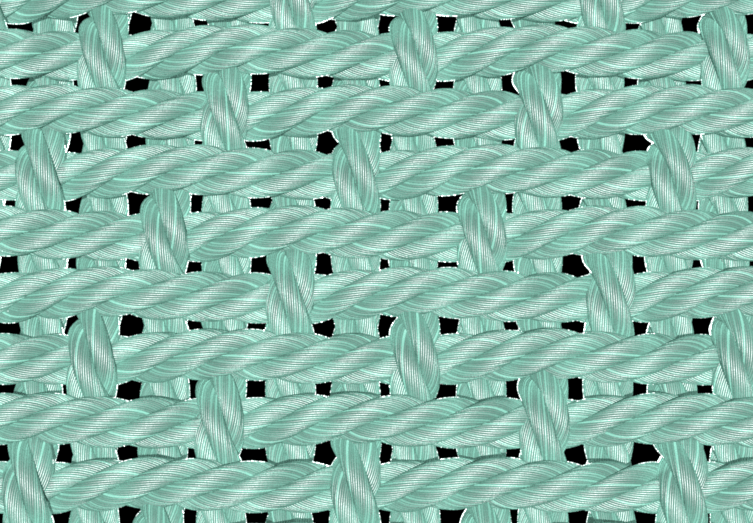} 
     \\ 
     \raisebox{0.0\textwidth}{\rotatebox{90}{Khung'15}}& 
     \includegraphics[width=0.16\textwidth, trim={0cm 4cm 0cm 0cm},clip]{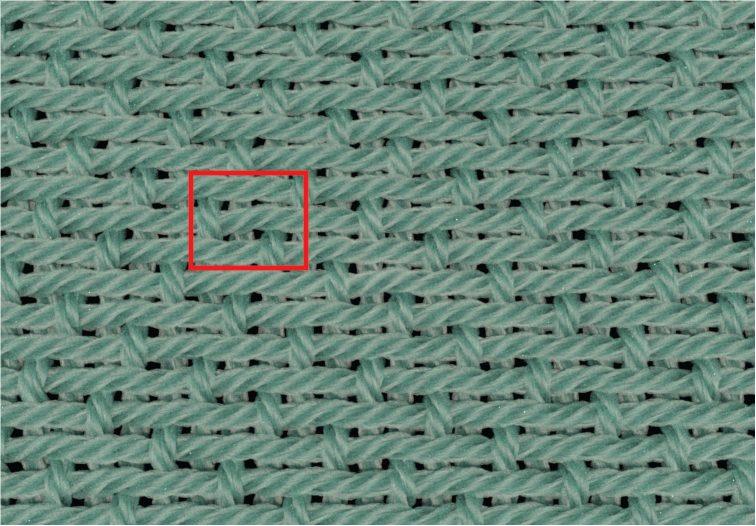} 
     &
     \includegraphics[width=0.16\textwidth, trim={0cm 4cm 0cm 0cm},clip]{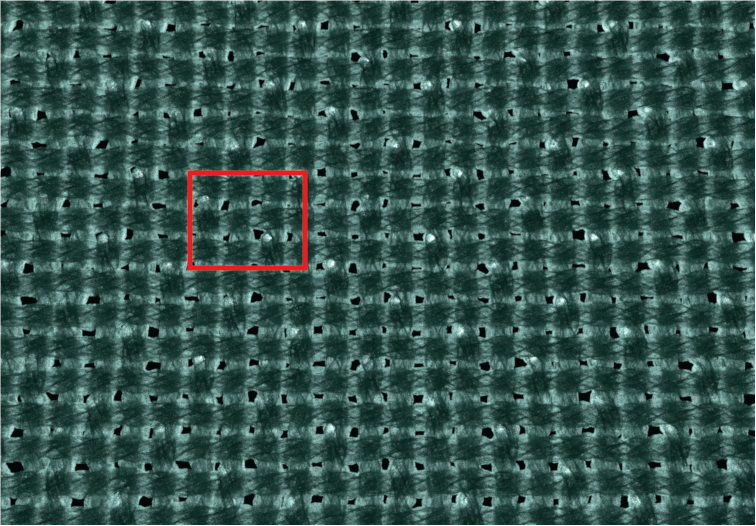} &
     \includegraphics[width=0.16\textwidth, trim={0cm 4cm 0cm 0cm},clip]{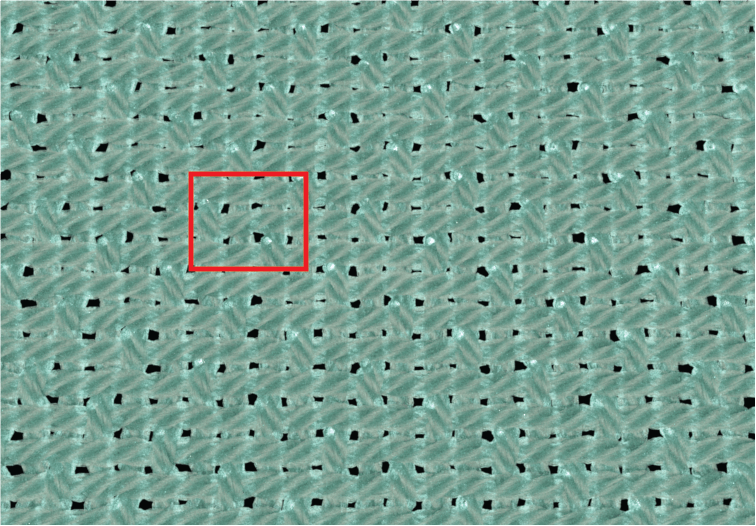} &
     \includegraphics[width=0.16\textwidth, trim={0cm 4cm 0cm 0cm},clip]{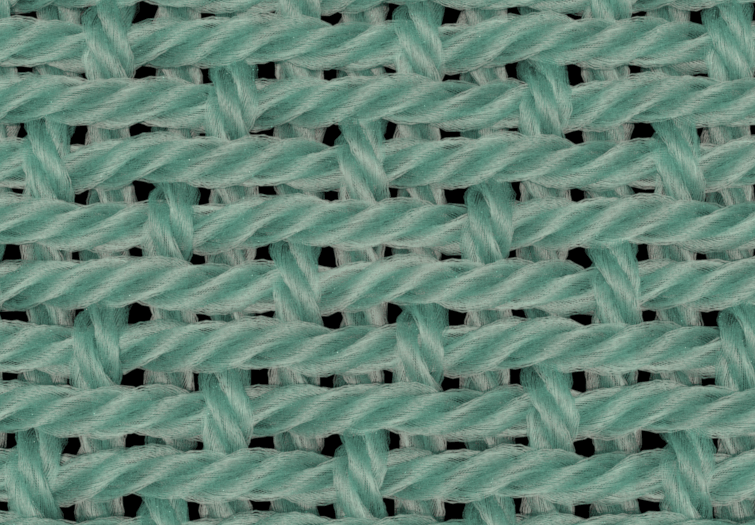} 
     &
     \includegraphics[width=0.16\textwidth, trim={0cm 4cm 0cm 0cm},clip]{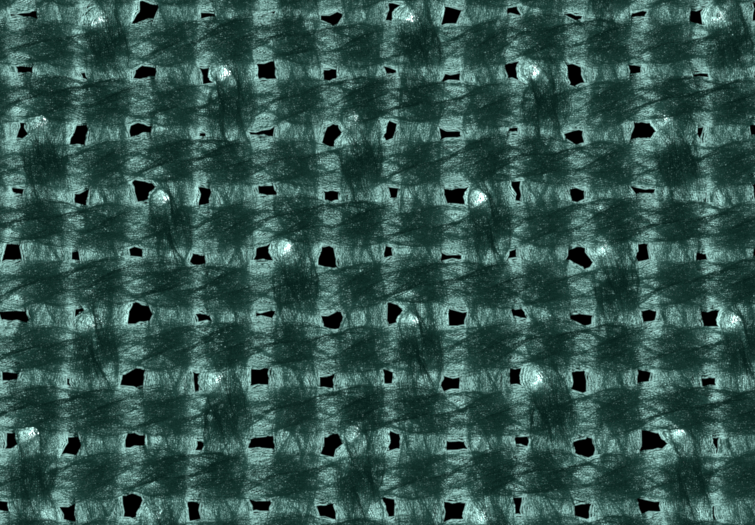} &
     \includegraphics[width=0.16\textwidth, trim={0cm 4cm 0cm 0cm},clip]{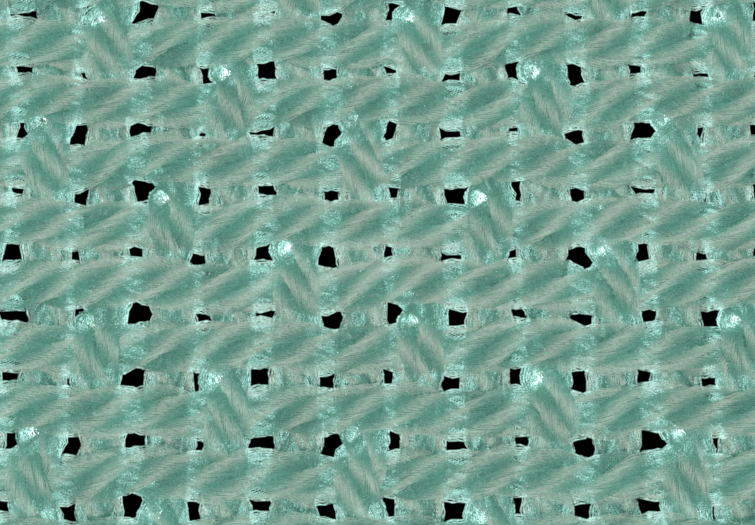} 
     \\ 
     & \multicolumn{3}{c}{-------------- Zoom-out --------------}
     & \multicolumn{3}{c}{-------------- Zoom-in --------------}
     \\
     \end{tabular}
    \caption{\label{fig_macro_comp_woven}{Comparison to previous fiber-based model.} Our model provides a close match to the reference ply-based model, more efficiently. A quantitative performance comparison can be found in Table \ref{table_performance}. Further comparisons are provided in the supplementary video. For distant renderings, our multi-scale solution further enhances efficiency. In comparison, the fiber-based model can achieve a similar match to the reference, but at a significantly higher computational cost as shown in zoomed-in rendering and the focus box in zoomed-out renderings. The woven samples are four- and five-ply yarns, respectively.}
\end{figure*}

\begin{figure*}[ht]
    \centering
    \setlength{\tabcolsep}{1pt}
    \begin{tabular}{cccccc}
         a) Near-field & b) Multi-scale & c) Far-field & a) Near-field & b) Multi-scale & c) Far-field 
     \\
     \includegraphics[width=0.16\textwidth,height=0.105\textwidth]{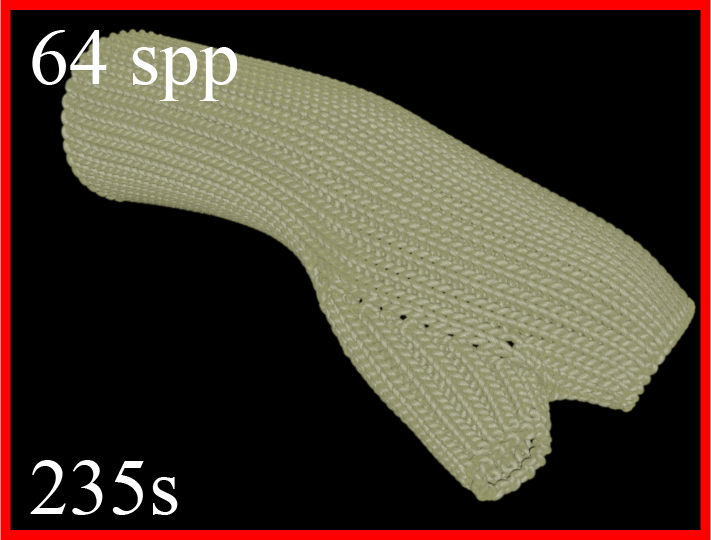} &
     \includegraphics[width=0.16\textwidth,height=0.105\textwidth]{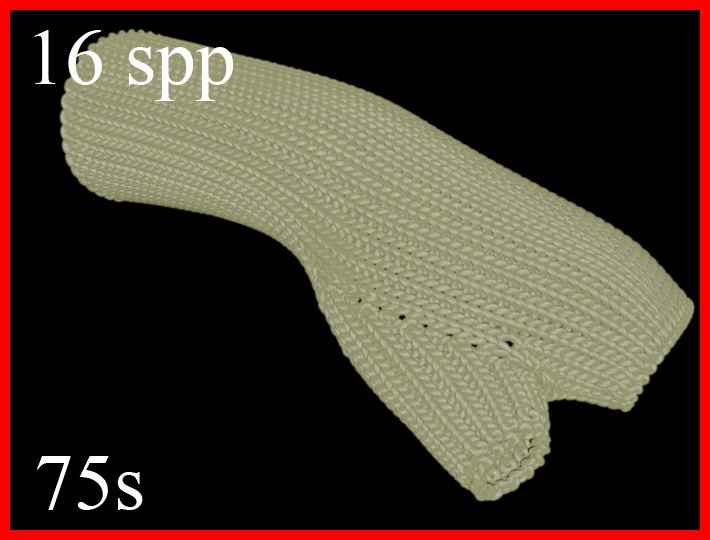} &
     \includegraphics[width=0.16\textwidth,height=0.105\textwidth]{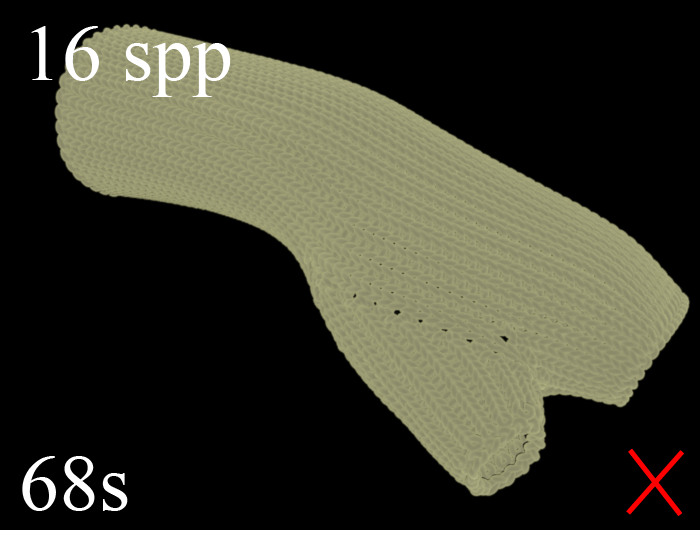} &
     \includegraphics[width=0.16\textwidth,height=0.105\textwidth]{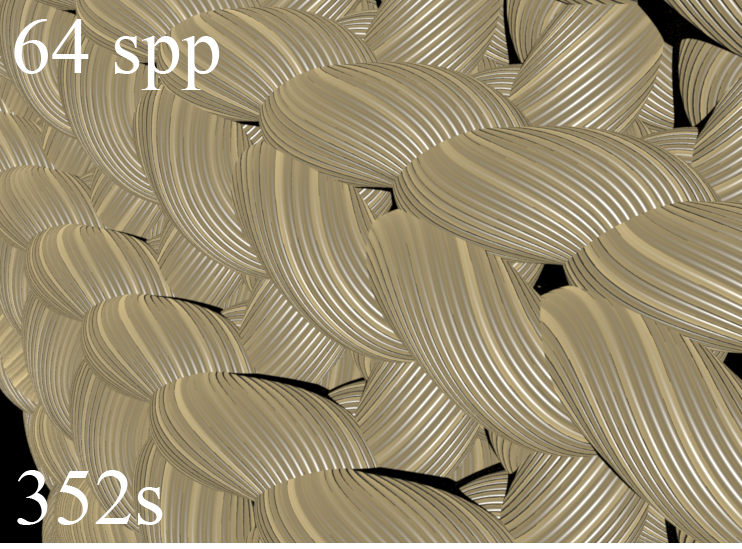} &
     \includegraphics[width=0.16\textwidth,height=0.105\textwidth]{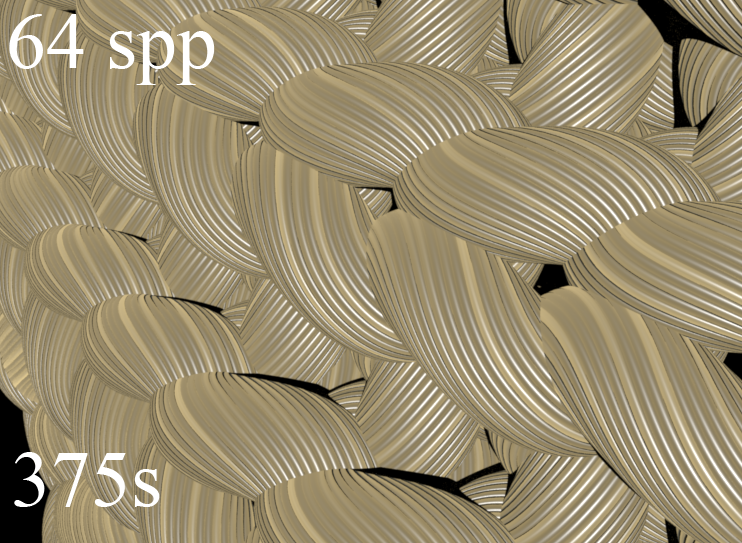} &
     \includegraphics[width=0.16\textwidth,height=0.105\textwidth]{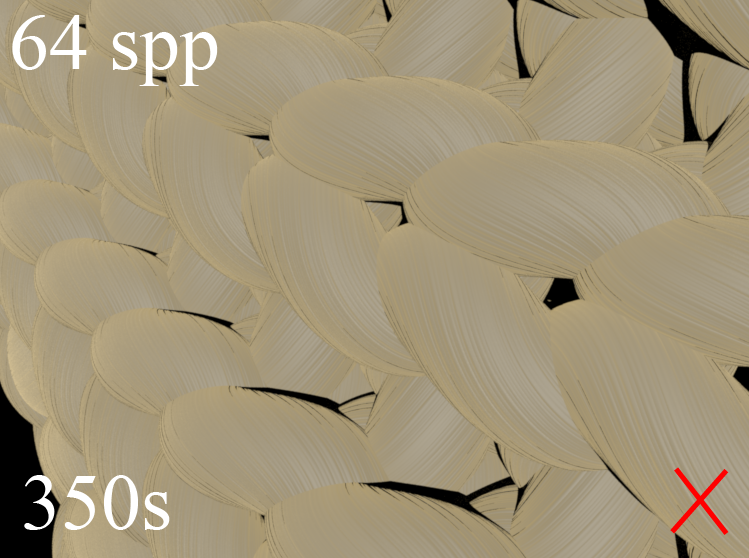} 
     \\
     \includegraphics[width=0.16\textwidth,height=0.105\textwidth]{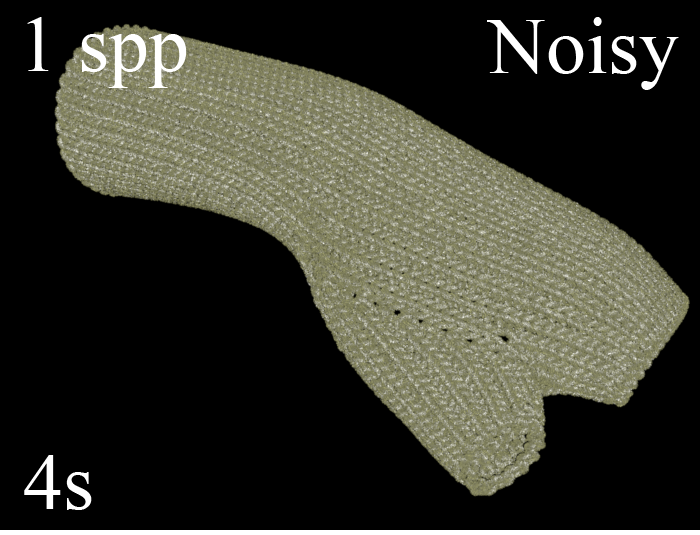} &
     \includegraphics[width=0.16\textwidth,height=0.105\textwidth]{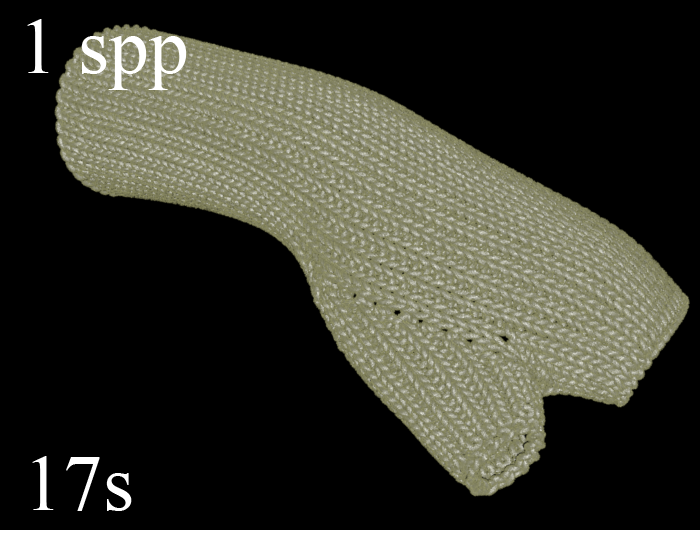} &
     \includegraphics[width=0.16\textwidth,height=0.105\textwidth]{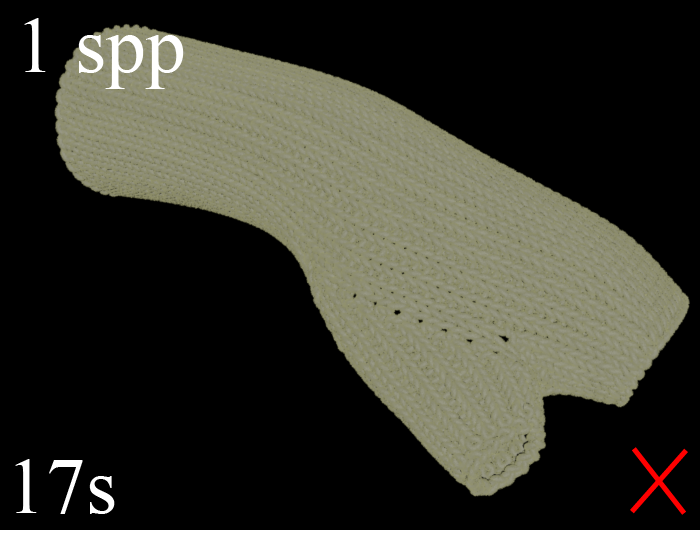} &
     \includegraphics[width=0.16\textwidth,height=0.105\textwidth]{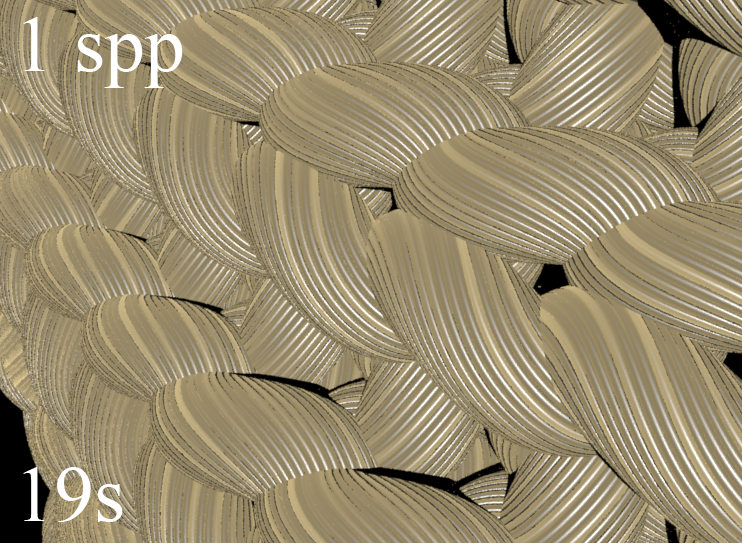} &
     \includegraphics[width=0.16\textwidth,height=0.105\textwidth]{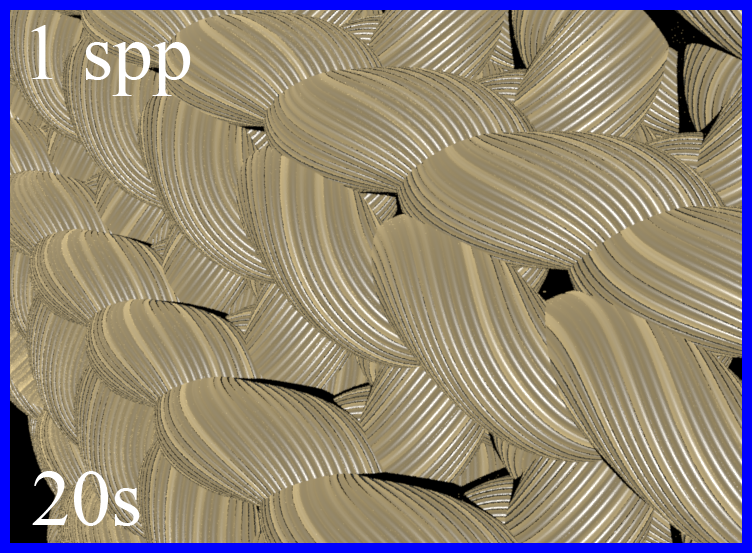} &
     \includegraphics[width=0.16\textwidth,height=0.105\textwidth]{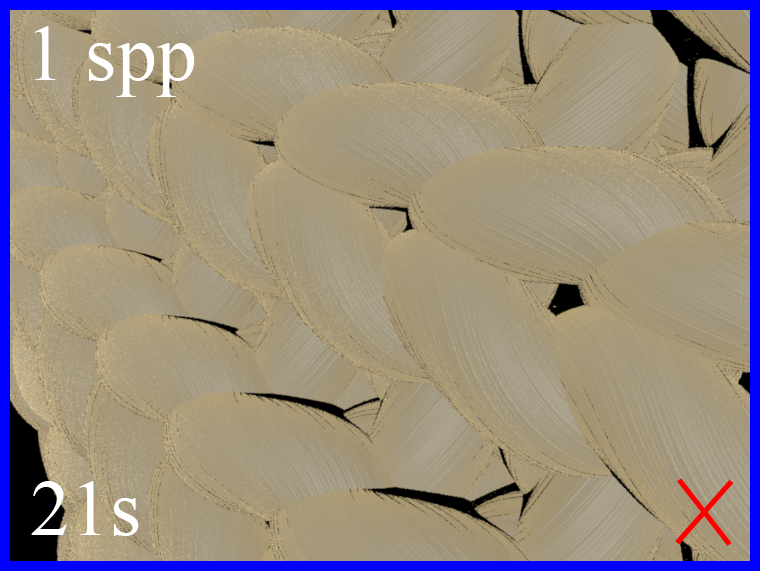} 
     \\
    \end{tabular}
    \caption{\label{fig_multires_com} {Multi-scale comparison.} Near-field mode captures fine details but becomes inefficient as the camera moves away. In the multi-scale mode, we smoothly transition between near-field and far-field renderings using an adaptive scheme based on pixel coverage, requiring fewer samples. Far-field mode efficiently captures macro-scale appearance when yarn width is smaller than pixel coverage. Comparing equal quality rendering (EQ) for a distant view (left 3 columns), our multi-scale model is around 2.4$\times$ faster than the near-field model. Similarly, for a close-up (right 3 columns), our multi-scale model offers better fiber details than the far-field model, in a similar time. Note in the third column the yarns are still larger than one pixel so we should not use the far-field method, and this is provided for comparison only, as indicated by a red cross. 
    }
\end{figure*}

\subsection{Multi-Scale Implementation}
\label{ssec_multi_detail}
To enable  multi-scale adaptation for BYSDF, we made modifications to the ray-sampler in Mitsuba 3. Instead of generating standard rays, we generate samples of the RayDifferential structure. The RayDifferential consists of the actual ray, along with two offset rays used for pixel coverage measurement. These offset rays provide different perspectives and allow us to capture a range of azimuthal offsets for the multi-scale feature. It is important to note that classical differentials correspond to the pixel footprint. In our case, we compute the intersected point when landing on a yarn and hence they might not be aligned with the azimuthal plane. To address this, we implemented custom ray differentials that ensure the range [$\azimuthalOffset_1$-$\azimuthalOffset_2$] corresponds to the actual azimuthal plane range within the pixel. This allows us to sample within this range adaptively based on the trajectories of pixels on the yarn geometries.

To determine the azimuthal offset range ($\azimuthalOffset_1$ and $\azimuthalOffset_2$), we utilize the ray-to-surface hit distance. By calculating the distances between the hit points after intersecting the ray differentials, we can define the range of azimuthal offset. $\azimuthalOffset_1$ corresponds to the distance between the original hit point at $\azimuthalOffset_0$ and its offset point in one direction, while $\azimuthalOffset_2$ corresponds to the distance between $\azimuthalOffset_0$ and its offset point in the opposite direction. In this scenario, the range of $\azimuthalOffset_1$ and $\azimuthalOffset_2$ is dependent on the yarn radius, and for simplicity, we normalized them to ensure a range of $(-1,1)$. These values use the origin ($o, o_1, o_2$) and direction ($d, d_1, d_2$) parameters as visualized in \figref{fig_multi_scale}.



\section{Results}
\label{sec_results}
\subsection{Setting}
In the following section, we present  rendering results obtained using our practical BYSDF and conduct a comparative analysis to existing approaches. The reference model in this paper is the ply-based model, which was evaluated against real photographs in the original paper. Consequently, we omit direct comparisons with actual photographs in this paper. We use the differentiable setup in Mitsuba 3 to match the set of parameters of our model to the reference. The details regarding parameter fitting can be found in the supplementary material. We also compare the BYSDF with the fiber-based model as well as Zhu et al. \cite{Zhu2023Hierarchical} which we refer to as the hierarchical yarn-based model. Our implementation of the geometric and appearance representation is integrated into the Mitsuba renderer \cite{Mitsuba3} as a custom integrator and material plugin. All scenes shown in the paper were rendered using a 4.50 GHz AMD Ryzen 9 7950X 16-core processor.

\subsection{Single yarns}
\subsubsection{Individual yarns}
In the first set of results (\figref{fig_micro_comp}), we compare our yarn-based method to the ply-based reference as well as the previous models under three different lighting configurations: front lighting only, back lighting only, and both front and back lighting. Our yarn-based model, like the reference, handles reflection and transmission separately, which allows it to capture the complex appearance of the yarn more accurately compared to the fiber-based model that considers all individual fiber interactions. We observe that the fiber-based model can match the appearance of the reference model reasonably well under front lighting conditions, but it struggles to match the appearance when the lighting is changed to the back or when both front and back lighting are present, highlighting the limitations of its complex geometry representation. Furthermore, our model offers more accurate back-lighting due to its proper transmission sampling, while the hierarchical yarn-based model results in an unwanted sharp silhouette from the plies in the back visible due to  transparency. This artifact arises because the refracted ray in their approach is employed without any sampling or perturbation; it is directly utilized as transmission, resulting in this visual discrepancy. Finally, our model demonstrates improved efficiency compared to the reference ply-based model, especially when dealing with yarns with multiple plies. 

\begin{table*}[ht]
\centering
\caption{Performance statistics. All rendering times are counted at equal quality. We give the time, memory required and number of bounces of our method (Ours), the ply-based model (Reference), the fiber-based model (FB) and hierarchical yarn-based model (HYB). \#plies means the number of plies in a yarn. Our method on average is  1.7 times faster than the Reference, 1.3 times faster than the hierarchical yarn-based model and about 6 times faster than the fiber-based model. Our method uses 200 kB more memory than the hierarchical yarn-based method as we additionally store ply-level shadow maps and fiber-level shadow, local tangent and local normal maps for the yarn cross-sections.}
    \setlength{\tabcolsep}{4pt}

\begin{tabular}{lccccccccccccc}
\toprule  
& &\multicolumn{4}{c}{Time (min)} & \multicolumn{4}{c}{Memory (MB)}& \multicolumn{4}{c}{\#Bounces}
\\\cmidrule(lr){3-6}\cmidrule(lr){7-10}\cmidrule(lr){11-14} scene & \#plies
         & Ours & Reference & FB&HYB & Ours & Reference & FB&HYB & Ours & Reference & FB&HYB\\\midrule
Fig. \ref{fig_micro_comp} single-ply & 1 & \important{0.1} & {0.1} & 2.5&0.2 & 0.06 & 0.06 & 17.3&\important{0.05} & \important{1} & 1 & 7.6&1 \\
Fig. \ref{fig_micro_comp} three-ply  & 3 & \important{0.1} & {0.3} & 2.2&0.2 & 0.06 & 0.18 & 31.5&\important{0.05} & \important{1} & 1.6 & 6.7&1 \\
Fig. \ref{fig_macro_comp_knit} knit zoom-out & 6 & \important{7.4} & {12.2} & -&10.2 & 9.1 & 53.2 & -&\important{8.9} & \important{3.6} & 5.1 & -&3.6 \\
Fig. \ref{fig_macro_comp_knit} knit zoom-in & 6 & \important{7.5} & {12.4} & -&10.3 & 9.1  & 53.2 & -&\important{8.9} & \important{3.7} & 5.3 & -&3.7 \\ 
Fig. \ref{fig_macro_comp_woven} basket zoom-out & 5 & \important{5.2} & {10.4} & {37.6}&- & \important{73.0} & 364.5 & 1520&- & \important{1.9} & 4 & 24.0&- \\
Fig. \ref{fig_macro_comp_woven} basket zoom-in  & 5 & \important{5.7} & {10.5} & 38.1&- & \important{73.0} & 364.5 & 1520&- & \important{1.7} & 4 & 22.7&- \\
Fig. \ref{fig_macro_comp_woven} satin zoom-out & 4 & \important{5.1} & {8.9} & {25.3}&- & \important{14.4} & 57.4 & 719&- & \important{1.6} & {3.9} &14.2&- \\
Fig. \ref{fig_macro_comp_woven} satin zoom-in  & 4 & \important{5.8} & {9.1} & {24.9}&- & \important{14.4} & 57.4 & 719&- & \important{1.6} & {3.8} &15.6&- \\

\bottomrule
\end{tabular}
\label{table_performance}
\end{table*}

\subsubsection{Woven and knitted samples} 
Furthermore, we conducted a comparison between our rendering results and the reference for woven and knitted samples to demonstrate that our yarn-based approach is capable of accurately reproducing the appearance of fabrics on a macro scale regardless of the manufacturing processes; the results are a reasonable match to the ply-based reference. \figref{fig_macro_comp_knit} presents a comparison between our BYSDF and the hierarchical yarn-based model. Our model captures the transmitted light accurately, but the other model has difficulty with back lighting because of the lack of sampling as explained in \figref{fig_micro_comp}. Besides, the overall appearance of its result is hard and does not reproduce the softness of the reference cloth. In \figref{fig_macro_comp_woven}, the fiber-based model reasonably matches the reference only if we enforce the parameters for each of the three setups separately. Our model, however, is only fitted for front- and back-lighting setups and automatically matches the front-and-back lighting scene with the reference due to decoupling of the reflection and transmission parameters. Our model is about 6 times faster due to the simplified geometry in comparison to the fiber-based model. The yarn geometries of all three samples were taken from the dataset by Leaf et al. \cite{Leaf2018Yarnsim} and the fiber curves were generated procedurally \cite{Zhao2016fitting} and shaded accordingly. Please refer to the Table \ref{table_performance} for a quantitative comparison.

\begin{figure*}[t]
    \centering
    \setlength{\tabcolsep}{1.5pt}
    \begin{tabular}[t]{lccl}
    \multirow{2}{*}[0.9in]{\includegraphics[width=0.32\linewidth, height=0.301\linewidth]{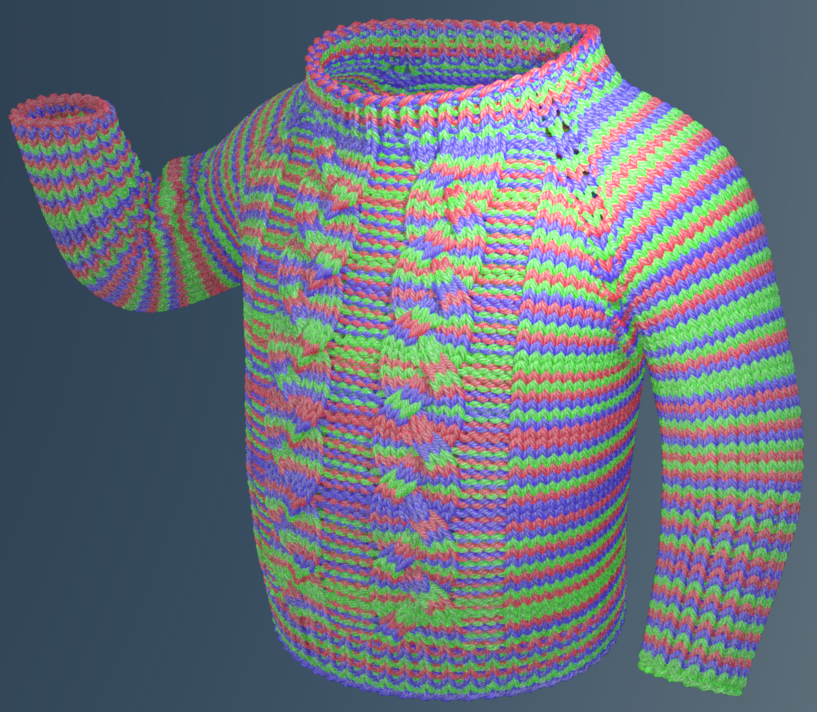}} & \includegraphics[width=0.16\linewidth]{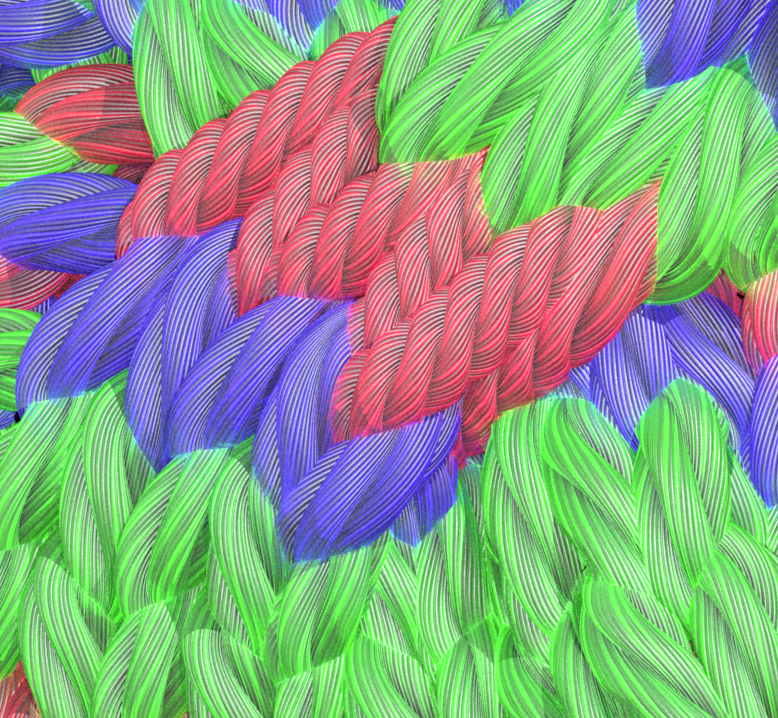} & \includegraphics[width=0.16\linewidth]{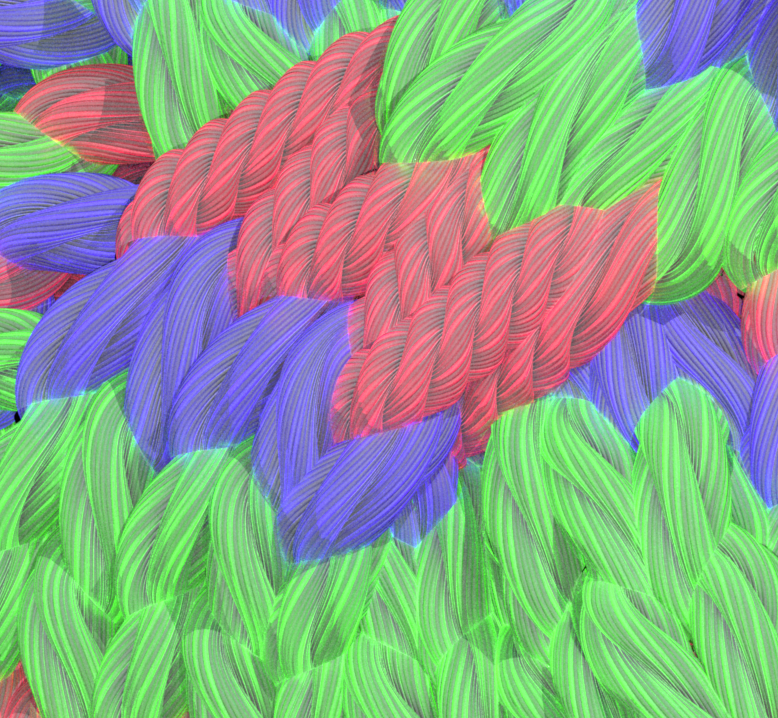}
    & \multirow{2}{*}[0.9in]{\includegraphics[width=0.32\linewidth, , height=0.301\linewidth]{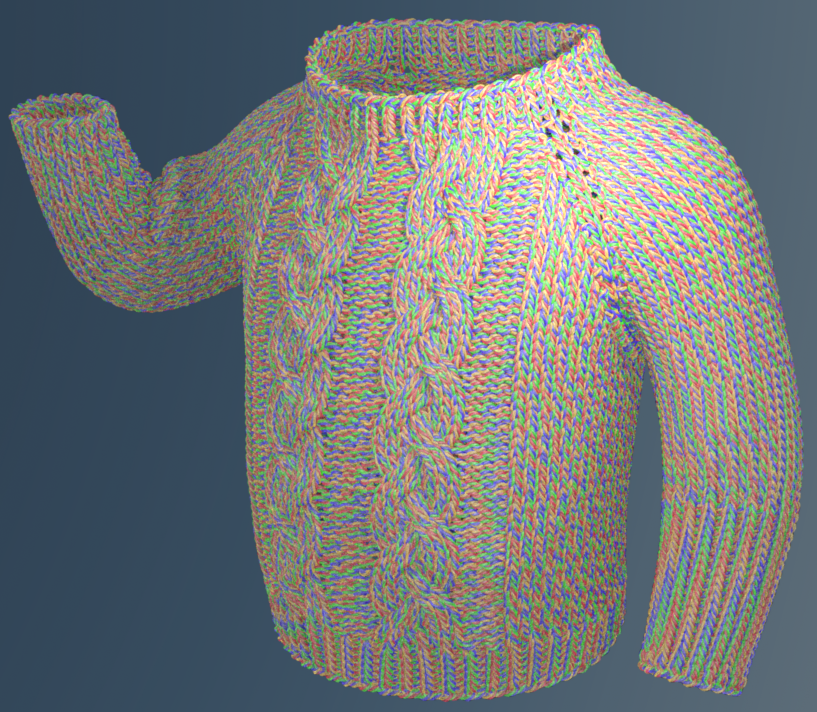}} \\
    & \includegraphics[width=0.16\linewidth]{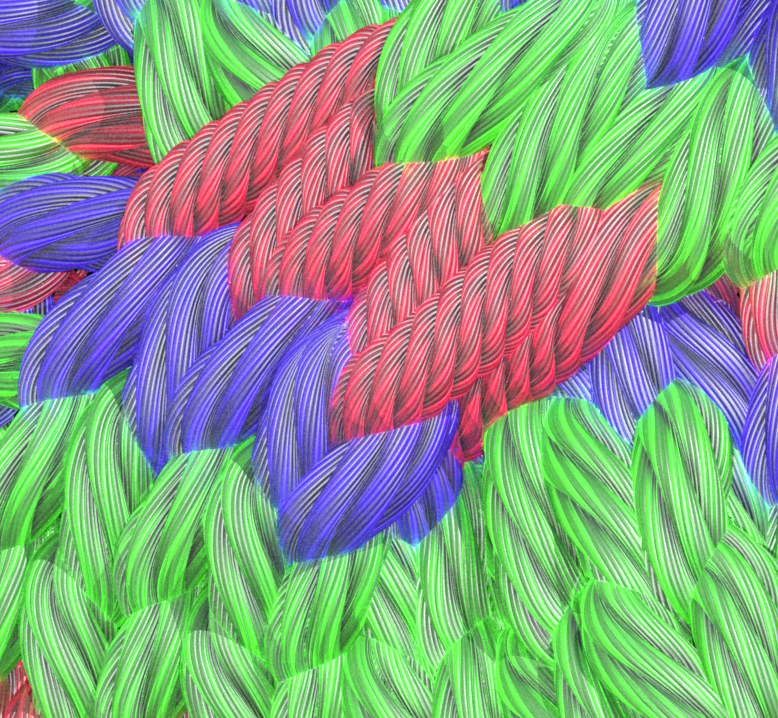} & \includegraphics[width=0.16\linewidth]{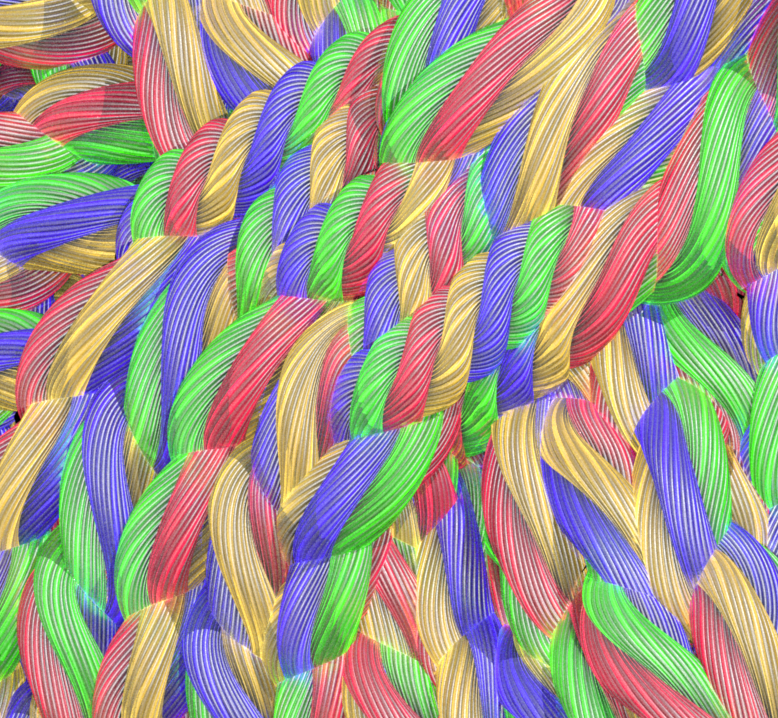} & \\
    \end{tabular}
    \put(-324,2){\color{white}\normalsize{8.1min}}
    \put(-242,2){\color{white}\normalsize{8.1min}}
    \put(-324,-75){\color{white}\normalsize{8.1min}}
    \put(-242,-75){\color{white}\normalsize{8.1min}}
    \put(-159,-75){\color{white}\normalsize{1.3min}}
    \put(-485,-75){\color{white}\normalsize{1.3min}}
     \caption{\label{fig_sweater} {Additional results.} Sweater model taken from the yarn dataset shaded using BYSDF at 128 spp. Left to right:,  a zoomed-out image with multiple colors of four-ply yarns, (row I) a zoomed-in image, (row I) yarns with  greater roughness, row(II) yarns with an increased number of plies (6), (row II) yarns with a different color for each ply, and corresponding zoomed-out image. The supplementary video showcases an animated multi-color sweater.
     }
     \label{rrm13}
\end{figure*}

\subsubsection{Multi-scale results} 
In \figref{fig_multires_com}, we illustrate the three different modes of our model. \figref{fig_multires_com} (a) illustrates the near-field mode, in which the model is optimized for rendering when the camera is close to the fabric; it  accurately captures the fine details of the fibers and ply-level geometry. However, as the camera moves away, rendering using the near-field model becomes noisy and inefficient.   To address this, we introduce the multi-scale mode---see \figref{fig_multires_com} (b)---which offers a smooth transition between  near-field and far-field rendering. This model is able to adaptively match the appearance of far-away rendering using a significantly smaller number of samples per pixel (spp) compared to the near-field mode. \figref{fig_multires_com} (c) shows the far-field mode, in which the model is integrated over the entire yarn, capturing the macro-scale appearance of the fabric with no fiber detail. This mode is suitable when the camera is far  from the fabric and whole pixels cover  more than the yarn width . The geometry of the glove scene is taken from the yarn dataset by Yuksel et al. \cite{Yukselyarns}.

Please note that the far-field appearance in both cases is incorrect because, even when zoomed far away (left), a yarn is still much thicker than a pixel. As a result, the actual range of $\azimuthalOffset$ is much smaller than the range of [-1, 1] used for far-field rendering. Therefore, the far-field appearance should not be expected to match the other modes. However, the far-field mode still enables a significant reduction in the number of ray samples compared to the near-field mode. In essence, when the fabric is viewed from close-up (right), our multi-scale method should provide rendering performance comparable to the near-field mode. This is achieved by dynamically determining the number of ray samples proportional to the azimuthal offset range ($\azimuthalOffset_2 - \azimuthalOffset_1$). On the other hand, when the fabric is viewed from far away, our method should be significantly more efficient than the near-field mode, as it allows for a substantial reduction in the number of required ray samples.

\subsubsection{Additional Results} 
We rendered the sweater model, taken from the yarn dataset by \cite{Yukselyarns} for varying geometry and appearance parameters under environment, area and point lighting. See Fig. \ref{rrm13}.

\begin{figure}[ht]
    \centering
    \setlength{\tabcolsep}{1pt}
    \begin{tabular}{cc}  
    Measurement & Ours \\
    \includegraphics[width=0.48\linewidth]{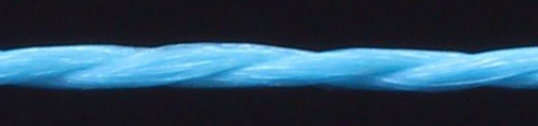} &
    \includegraphics[width=0.48\linewidth]{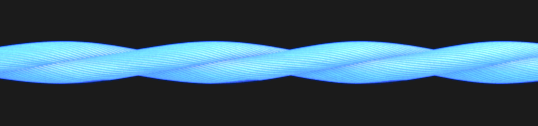} \\
    \includegraphics[width=0.48\linewidth]{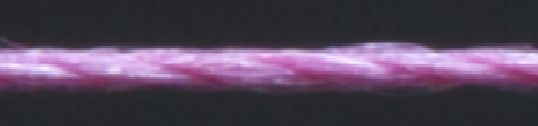} & 
    \includegraphics[width=0.48\linewidth]{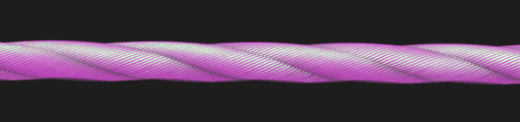}
    \end{tabular}
    \caption{\label{fig:comp_real}
 {Comparison to real samples} using real fiber data for rayon 2-ply and silk 3-ply. Our model produces a close match in appearance to the measurement lit by an area and a constant light.}
\end{figure}

We present results generated using real fiber data provided by Zhao et al. \cite{Zhao2016fitting} (Fig. \ref{fig:comp_real}). Notice that in the real data, the plies stretch and compress along the yarn (especially evident in the 2-ply case) while our method represents the plies implicitly using a fixed radius.  

\begin{figure}[ht]
    \centering
    \setlength{\tabcolsep}{1pt}
    \begin{tabular}{cccc}
    Lamp off & Twill & Satin & Rotated satin \\
    \includegraphics[width=0.24\linewidth]{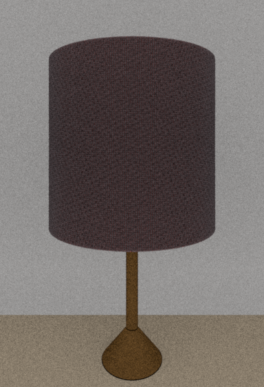} & \includegraphics[width=0.24\linewidth]{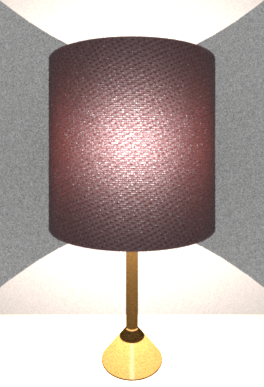} & \includegraphics[width=0.24\linewidth]{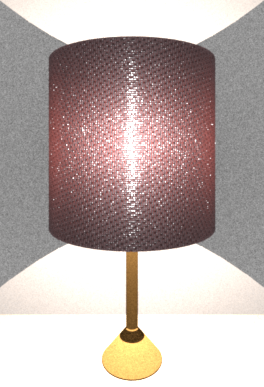} &
    \includegraphics[width=0.24\linewidth]{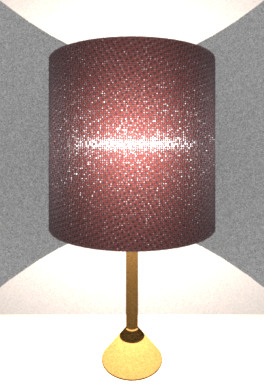} \\
    \end{tabular}
    \caption{Additional results. Lamp model for twill and satin pattern under constant and point light.}
    \label{fig:lamp}
\end{figure}

We also rendered a simple lamp with satin and twill patterns under constant and point lighting to show the anisotropic appearance of fabrics. See Fig. \ref{fig:lamp}.  The twill pattern exhibits a round highlight  whereas the satin exhibits an elongated highlight as shown in previous works \cite{montazeri2020practical}.

\section{Discussion and Conclusions}
\label{sec_conclusion}
\subsection{Limitations and Future Work} 
Our model, while efficiently adding fiber texture using 1D texture maps, has room for improvement in seamlessly integrating fiber details directly into the yarn cross-section, eliminating the need for texture maps. Besides, the loss of energy due to  approximation of  multiple scattering results in the same limitation as previous works. Additionally, capturing the appearance of flyaways, a common feature in real fabrics, is a potential area for enhancement in our framework. For future work, we also aim to model fabrics like velvet, which presents unique challenges due to its protruding fiber bundles.

\mypara{Conclusions} 
Our paper  introduces BYSDF, an effective yarn-based model that excels in rendering the intricate appearance of cloth, particularly in close-up views. It simplifies geometry, thereby reducing ray tracing costs, and avoids complex light transport between plies and fibers, resulting in a need for fewer rays. The model successfully transitions from near-field to far-field rendering, integrating geometry and appearance seamlessly as  camera distance changes.

BYSDF builds upon the ply-based model, aggregating shading at the yarn level to capture the implicit geometries of ply and fiber. We compensate for the lack of explicit geometry with realistic shadowing effects, enhancing performance and fidelity. Our model achieves faster rendering speeds and lower memory usage for near-field renderings, and our multi-scale solution further optimizes distant rendering.

\appendix


\subsection*{Acknowledgements}
We would like to thank  the reviewers as well as Wētā rendering team for their valuable inputs. This research was partially funded a  University of Manchester Dean's Award.



\subsection*{Declaration of competing interest}

The authors have no competing interests to declare that are relevant to the
content of this article. 

\bibliographystyle{CVMbib}
\bibliography{main}

\subsection*{Author biography}

\begin{biography}
[author_1]{Apoorv Khattar} is currently pursuing his Ph.D. at the University of Manchester under the supervision of Dr. Zahra Montazeri and is a recipient of a Dean's Doctoral Scholarship Award. His research interest is realistic cloth appearance modeling. He is undertaking  a research internship at Wētā Digital in 2024. He gained a bachelor's degree from Indraprastha Institute of Information Technology (IIIT), Delhi in 2020.
\end{biography}
\begin{biography}
[author_3]{Ling-Qi Yan} is an Assistant Professor of Computer Science at UC Santa Barbara, a co-director of the MIRAGE Lab, and affiliated faculty in the Four Eyes Lab. He strives to achieve photo-realistic rendering. His research focuses on physically-based rendering at both micro and macro scales, encompassing areas such as appearance modeling and representation, physical light transport theory, and real-time ray tracing practice. His  contributions have been recognized through various accolades, including the SIGGRAPH 2019 Outstanding Doctoral Dissertation Award, a SIGGRAPH 2022 Best Paper Honorable Mention and an EGSR 2023 Best Paper Award.
\end{biography}


\begin{biography}
[author_2]{Zahra Montazeri} is a lecturer (assistant professor) at the University of Manchester in the Department of Computer Science. Her field of research is rendering and appearance modeling for complex materials such as cloth. She has worked as a research consultant at Disney Research and received a  movie credit for Star Wars: The Mandalorian; her research  was used in the production of Avatar, The Way of Water. Before joining academia, she worked in R\&D at Industrial Light \& Magic  and interned at Pixar Animation Studios, DreamWorks Animation and Luxion. She holds a Ph.D. from the University of California, Irvine, and was awarded a B.Sc. by  Sharif University of Technology, Iran in 2015.
\end{biography}


\subsection*{Electronic Supplementary Material}
An accompanying video shows some of our results. It comprises  the following scenes:
\begin{enumerate}
    \item 5-ply Woven Basket: comparison to the reference with directional lighting and camera transition between close-up and far-view.
    \item 1-ply Glove: comparison to the reference with a moving light.
    \item 1-ply Glove: comparison to the reference with  rotating geometry.
    \item 1-ply Glove: multi-scale comparison to the reference with directional lighting and a camera transition between close-up and distant-view.
    \item 3-ply multi-color sweater: with environment, area and sharp point lighting, and a camera transition between close-up and distant-view.
\end{enumerate}

\end{document}